\documentclass[sigconf,nonacm]{acmart}

\AtBeginDocument{%
  }

\usepackage{subcaption}
\usepackage{booktabs}
\usepackage{multirow}
\usepackage{ragged2e}
\usepackage{array}
\usepackage{longtable}
\usepackage{listings}
\usepackage{xcolor}
\usepackage{amsmath}
\usepackage{tikz}
\usetikzlibrary{arrows.meta,calc,fit,positioning}

\renewcommand\footnotetextcopyrightpermission[1]{}
\newif\iflayoutpreview
\ifdefined\LAYOUTPREVIEW
  \layoutpreviewtrue
\else
  \layoutpreviewfalse
\fi

\title{SocialCoach: Personalized Social Skill Learning with Agentic Tutoring and Practice}

\author{Tianfu Wang}
\affiliation{%
  \institution{HKUST (GZ)}
  \city{Guangzhou}
  \country{China}
}
\email{twang566@connect.hkust-gz.edu.cn}

\author{Max Xiong}
\affiliation{%
  \institution{Duke University}
  \city{Durham}
  \country{United States}
}
\email{mx71@duke.edu}

\author{Yuxuan Lei}
\affiliation{%
  \institution{Microsoft}
  \city{Beijing}
  \country{China}
}
\email{t-yuxuanlei@microsoft.com}

\author{Jianxun Lian}
\affiliation{%
  \institution{Microsoft Research Asia}
  \city{Beijing}
  \country{China}
}
\email{jianxun.lian@outlook.com}

\author{Hongyuan Zhu}
\affiliation{%
  \institution{Microsoft}
  \city{Beijing}
  \country{China}
}
\email{hongyuanzhu@microsoft.com}

\author{Zhengyu Hu}
\affiliation{%
  \institution{HKUST (GZ)}
  \city{Guangzhou}
  \country{China}
}
\email{v-huzhengyu@microsoft.com}

\author{Linxiao Gong}
\affiliation{%
  \institution{HKUST (GZ)}
  \city{Guangzhou}
  \country{China}
}
\email{Igong265@connect.hkust-gz.edu.cn}

\author{Dapeng Hu}
\affiliation{%
  \institution{Microsoft}
  \city{Beijing}
  \country{China}
}
\email{dapenghu@microsoft.com}

\author{Xiaofang Li}
\affiliation{%
  \institution{Microsoft}
  \city{Beijing}
  \country{China}
}
\email{xiaofangli@microsoft.com}

\author{Peiting Tsai}
\affiliation{%
  \institution{Microsoft}
  \city{Beijing}
  \country{China}
}
\email{peitingtsai@microsoft.com}

\author{Nicholas Jing Yuan}
\affiliation{%
  \institution{Microsoft}
  \city{Beijing}
  \country{China}
}
\email{nicholas.yuan@microsoft.com}

\author{Qi Zhang}
\affiliation{%
  \institution{Microsoft}
  \city{Beijing}
  \country{China}
}
\email{zhang.qi@microsoft.com}

\renewcommand{\shortauthors}{Tianfu Wang et al.}

\begin{document}

\begin{abstract}
Social skills such as negotiation and leadership are crucial for personal and professional success in today's interconnected world. However, scalable and effective training remains a significant challenge due to the scarcity of expert coaching. In this work, we introduce SocialCoach, an LLM-powered agentic tutoring system for personalized social skill learning. SocialCoach constructs a theory-to-practice corpus of traceable strategies, cases, and practice scenarios, and uses this corpus for both scheduling and reflective tutoring.
We formulate social practice personalization as cold-start, retrieval-constrained sequential practice scheduling. Given a learner profile, simulated proficiency state, and observed practice history, a policy produces structured prescriptions that are realized through corpus retrieval. To enhance scheduling effectiveness, we optimize complete pathways with trajectory-level GRPO using rubric-judge-based pairwise preferences.
Additionally, we instantiate the scheduling approach in a deployed platform with goal-driven practice and knowledge-grounded reflective tutoring. Finally, in the synthetic cold-start setting, experiment results show that SocialCoach achieves higher pathway-quality ratings than baselines in scheduling and tutoring quality. We also conduct human studies to demonstrate its usefulness for real-world social skill learning.
\end{abstract}



\maketitle

\section{Introduction}

\begin{figure}
    \centering
    \includegraphics[width=0.98\linewidth]{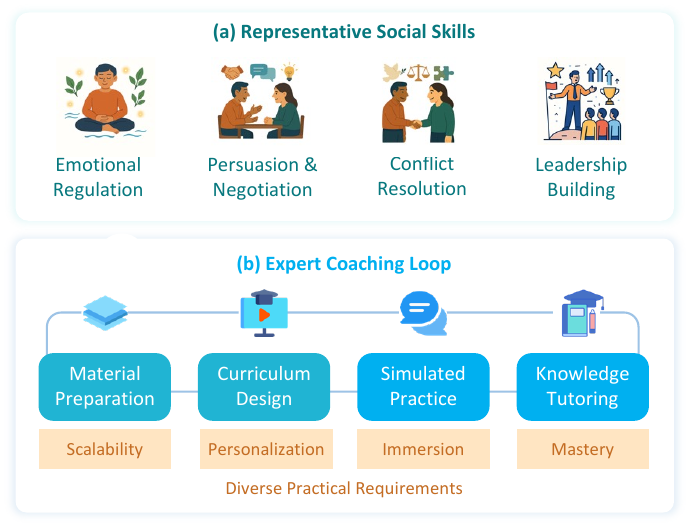}
    \vspace{-10pt}
    \caption{The social skills training landscape. (a) Examples of critical social skills. (b) The typical expert coaching loop, highlighting diverse practical requirements.}
    \label{fig:social-skill-coaching}
    \vspace{-15pt}
\end{figure}

In this increasingly interconnected and collaborative global landscape, proficiency in complex social skills is critical for both personal well-being and professional success. As illustrated in Figure~\ref{fig:social-skill-coaching}(a), social skills, such as negotiation, leadership, and conflict resolution, profoundly shape an individual's ability to navigate interpersonal environments~\cite{sst-dean2017soft}. These competencies not only underpin effective interpersonal relationships but also serve as significant drivers of economic value. Yet, despite their undeniable importance, the effective acquisition of these skills remains a critical gap for most learners worldwide.
Traditionally, expert coaching is a proven pathway for social skill development. As illustrated in Figure~\ref{fig:social-skill-coaching}(b), this educational loop consists of several key stages, including the knowledge preparation for specialized skills, tailored curriculum design, scenario simulation for practice and providing constructive guidance. However, effectively orchestrating these stages requires substantial human expertise and resources, which limits the scalability and accessibility of high-quality coaching~\cite{sst-ovink2011more}.

The advent of Artificial Intelligence (AI) presents a transformative opportunity to address these limitations.
However, early AI-driven methods often selectively focus on isolated educational tasks such as social dialogue generation or knowledge tutoring, and have limited generalizability over various topics.
Large Language Models (LLMs) have demonstrated a remarkable capacity for both vast knowledge integration and context-aware dialogue generation. This is crucial to distinguish their application in declarative knowledge acquisition from the experiential, practice-driven nature of social skill development \cite{social-arxiv-2024-llm-training}.
Several works have begun to leverage LLM for social conversational practice \cite{social-aaai-2025-social-tutor,social-chi-2025-gptcoach} and to develop frameworks for specific competencies like conflict resolution \cite{social-chi-2025-conflict-resolution} and empathy \cite{social-chi-2025-scaffolding-empathy}.
Despite these advances, existing approaches typically target individual coaching stages or specific social skills and do not fully meet the practical requirements. Consequently, a significant gap persists in the absence of a unified social skill development framework that can support comprehensive social skills, adaptive scenario practice, and pedagogically grounded tutoring. This highlights a clear need for a holistic framework that enables scalable, personalized training experiences.

To build such a framework, three fundamental data science challenges must be addressed~\cite{sst-jones2017social}:
(a) \textit{Scalable Knowledge Discovery and Structuring}.
To ensure pedagogical rigor, a trustworthy knowledge base must be grounded in verifiable, expert-endorsed principles.
However, while LLM-generated content risks unverifiability and hallucination, expert knowledge in social coaching is predominantly unstructured and tacit, often residing in practitioner experience or scattered sources.
Thus, a core challenge lies in automatically extracting, codifying, and structuring validated expertise from authoritative sources at scale.
(b) \textit{Personalized Practice Curricula Scheduling}.
Real-world learners differ in their backgrounds, goals, target skills, and prior practice. Static curricula cannot account for these differences, while unconstrained generation may request scenarios that do not exist in the available corpus. The scheduling problem is therefore both personalized and retrieval-constrained: each prescription must fit the learner while remaining realizable as a traceable practice scenario.
(c) \textit{Diagnostic Coaching Scaffolding Development}.
Effective tutoring aims to promote learner autonomy in real-world social skill application. This requires accurately assessing the knowledge deficits in practices and scaffolding to bridge the "knowing–doing" gap. Besides, beyond prescriptive feedback alone, the system must enable guided reflection by linking practice to theory and diagnosing skill gaps. Delivering such diagnostic, reflective guidance remains an underexplored challenge.

To bridge these critical gaps, we introduce \textbf{SocialCoach}, an agentic tutoring and practice system for personalized social skill learning.
First, to ensure educational validity at scale, we develop a structured framework for organizing social skill knowledge, grounded in a theory-to-practice paradigm and enriched through multi-faceted semantic associations. An automated pipeline transforms unstructured expert books into a coherent corpus of strategic theories, illustrative cases, and practical scenarios. This corpus provides a traceable practice space for both personalized scheduling and knowledge-grounded tutoring.
Second, at the core of personalization, we formulate next-practice selection as a cold-start, retrieval-constrained scheduling problem. We develop a prescription--retrieval--adaptation policy that conditions on the learner's profile, simulated proficiency state, observed practice history, and available corpus. An expected proficiency estimator generates bounded skill-level transitions for each selected scenario to simulate state transitions. We optimize the scheduling policy for complete practice pathways using pairwise judge-based GRPO training.
Finally, to close the practice and tutoring loop, we introduce a pedagogically grounded coaching process that tightly integrates interactive practice with metacognitive scaffolding. After a learner engages in a goal-driven simulated scenario, the system performs an evidence-linked assessment of observed behavior and provides tailored, knowledge-grounded guidance.

In experiments, we evaluate corpus quality, pathway quality, tutoring quality, and perceived usefulness. Automated evaluation assigns higher quality ratings to SocialCoach than to compared baselines, which is well-aligned with experts.
Furthermore, the user study reveals a better learning experience with SocialCoach.

We highlight our main contributions as follows.

\begin{itemize}
    \item We introduce SocialCoach, an agentic tutoring and practice system for personalized social skill learning, and formulate its central personalization problem as cold-start, retrieval-constrained sequential practice scheduling.

    \item We establish a theory-to-practice social knowledge framework with multi-faceted semantic categorization. The resulting corpus provides a structured practice space for scheduling and knowledge-grounded tutoring.
    \item To address cold-start personalization, we propose a retrieval-constrained learning pathway scheduling algorithm that formulates practice planning as a prescription--retrieval--adaptation process. It also leverages the pairwise pathway GRPO to optimize the prescription policy.


    \item We instantiate SocialCoach in a deployed platform and evaluate it through corpus analysis, automated evaluation, and user study. Results show its effectiveness on corpus quality, tutoring quality, and user learning experience.
\end{itemize}

\vspace{-7pt}
\section{Related Work}

\begin{figure*}[!ht]
    \centering
    \input{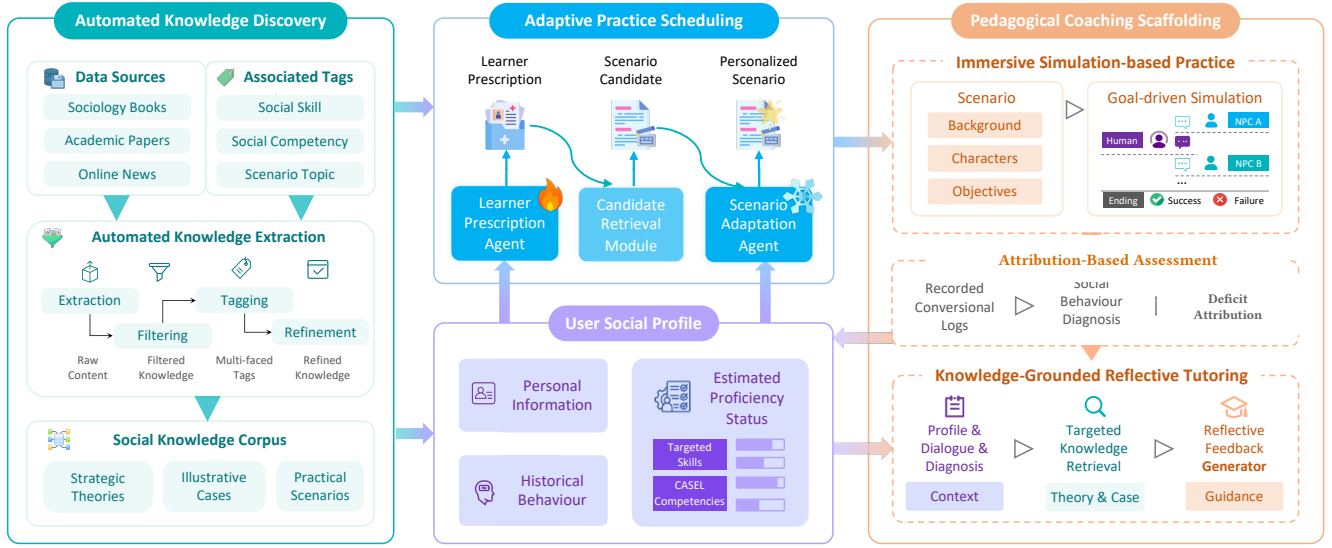}
    \vspace{-8pt}
    \caption{An overview of the SocialCoach Framework for personalized social skill training at scale.}
    \label{fig:social-coach-framework}
    \vspace{-12pt}
\end{figure*}

\noindent
\textbf{LLM-powered Tutoring Systems.}
LLMs are increasingly being leveraged to enhance Intelligent Tutoring Systems (ITS), which serve as conversational partners, moving beyond the static, content-centric traditional platforms~\cite{education-arxiv-2025-survey-agent,jws-2025-education-ai}.
Prior work has explored specialized functions within ITS, including learner profiling~\cite{education-ijcai-2025-coderagent} and performance assessment~\cite{knowledge-arxiv-2025-wise}. Also, there are some studies on enhancing pedagogical strategies of chatbot tutors like Socratic guidance~\cite{education-neurips-2024-socraticlm-personalized-teaching,education-arxiv-2025-teaching-problem-solving}.
Furthermore, several multi-agent architectures~\cite{education-www-2025-gen-mentor,education-emnlp-2025-conversational-scale} have been proposed to orchestrate complementary pedagogical roles, where specialized agents collaborate on tasks such as content creation, feedback, and evaluation.
Despite these advances, most LLM-powered tutors remain oriented toward knowledge transmission rather than practice-based development. However, social skill learning emphasizes iterative practice, contextual sensitivity, and behavioral transfer. Thus, a critical gap remains in developing scalable, personalized, and pedagogically grounded LLM-driven tutoring and practice systems tailored to social skill learning~\cite{social-arxiv-2024-llm-training}.



\noindent \textbf{Intelligent Social Emotional Learning.}
Social–emotional learning (SEL) focuses on helping learners cultivate positive attitudes, self-awareness, and healthy interpersonal relationships~\cite{sst-book-2025-sel-casel}. Early AI-driven efforts in this space emphasized conversational support, real-time feedback, and scaffolds for socio-emotional growth~\cite{sst-jritl-2024-survey}.
With the rise of LLMs, the field has shifted toward adaptive, role-playing–based systems that can target specific skills with greater nuance~\cite{social-arxiv-2024-llm-training}.
Recent LLM-based approaches have demonstrated promise in skills such as conflict resolution~\cite{social-chi-2025-conflict-resolution}, empathy coaching~\cite{social-chi-2025-scaffolding-empathy}, and stress self-regulation~\cite{social-chi-2025-stress}
However, despite these advances, current systems are typically scoped to narrow skill areas and face persistent challenges around knowledge traceability and pedagogical reliability. This fragmentation limits the scalability and pedagogical grounding of learning experiences of social skill learning. It also hinders cross-context skill transfer, creating a “knowing–doing” gap between understanding and practical use.




\noindent
\textbf{Personalization in Educational System.}
Personalized learning experience is vital for improving efficiency and engagement as learners progress~\cite{education-spm-2021-personalized-education}. Particularly, designing adaptive learning pathways can better match individual needs than traditional or rule-based curricula.
To improve path personalization, prior work~\cite{education-www-2024-mooc,education-ls-2025-planglow} trains models for specific scenarios using supervised learning on well-labeled data. However, such data are often proprietary or scarce in many educational scenarios. To address this cold-start problem, recent studies~\cite{education-chi-2020-rl-educational-activities,education-kdd-2024-knowledge-state-distillation} use simulated environments and reinforcement learning to optimize learning-path policies. These simulators are typically designed for structured domains such as mathematics and computer science~\cite{education-chi-2020-rl-educational-activities}; social-skill practice instead requires sequencing open-ended scenarios under learner and corpus constraints. We therefore use a bounded, scenario-conditioned proficiency transition as a cold-start planning proxy, rather than asking an LLM learner to generate and then self-assess an entire practice response.
Outside educational scheduling, profile-conditioned retrieval uses persona descriptions as queries over external corpora~\cite{retrieval-emnlp-2023-lapdog}, personalized query expansion incorporates user-specific signals before retrieval~\cite{retrieval-aaai-2026-personalize-before-retrieve}, and history-aware retrieval reformulates queries using relevant prior turns~\cite{retrieval-findings-2024-haconvdr}. These mechanisms motivate our matched retrieval controls, but do not themselves optimize multi-session practice pathways.

\vspace{-7pt}
\section{Problem Definition}

Let $\mathcal{S}$ denote the available social skills and $\mathcal{K}_s$ the corpus of traceable practice scenarios. A synthetic learner is represented by profile information $U$, stated goals $G \subseteq \mathcal{S}$, an initial proficiency vector $U_p^0\in[1,5]^{|\mathcal{S}|}$, and an append-only history $H_t$ of realized practice and retrieval events. The initial scores are part of profile generation rather than measurements of real users. From $\mathcal{K}_s$, $U_p^t$, and $H_t$, the system derives the available practice candidates.


At session $t$, the scheduler observes the state $s_t = (U,G,U_p^t,H_t)$ and produces a structured prescription $a_t$. The retriever realizes this prescription as a corpus scenario $p_t$. Given $p_t$, the tutoring system guides the learning. The planning environment then updates the learner state and practice history:
$
s_{t+1}=\mathcal{T}(s_t,a_t,p_t).
$
A pathway is sequential because each selected practice changes the state observed by subsequent scheduling decisions. We seek a policy that produces personalized, coherent, and progression-aware pathways under corpus and horizon constraints in cold-start settings.

\vspace{-7pt}
\section{The SocialCoach Framework}
To achieve personalized social skill learning at scale, we introduce \textbf{SocialCoach}, an agentic tutoring and practice system illustrated in Figure~\ref{fig:social-coach-framework}.
SocialCoach holistically integrates several critical components.
Firstly, it automatically constructs a comprehensive and pedagogically-grounded social skills knowledge corpus from diverse, unstructured sources using LLMs.
Second, it maintains a user profile and a factual history of completed practices and retrieval events.
Additionally, SocialCoach employs an adaptive practice scheduling policy trained with reinforcement learning to prepare scenario sequences tailored to individual goals. For each practice, it delivers pedagogically grounded coaching through immersive practice, behavior-grounded assessment, and personalized feedback.
We elaborate on each of these components below.

\vspace{-5pt}
\subsection{Automated Knowledge Discovery}
\label{section:automated-knowledge-discovery}
Effective social skill training depends on the availability of high-quality, context-rich pedagogical content that reflects real-world interpersonal challenges. However, expert knowledge in this domain is often tacit and dispersed across unstructured sources such as books, articles, and media~\cite{sst-berj-2011-socially-skilled-teacher}.
Manual curation of such knowledge is labor-intensive and difficult to scale, while direct LLM-based content generation risks unverifiability and hallucination~\cite{llm-www-2025-hallucination}. These methods compromise knowledge scalability or pedagogical reliability.
To overcome these challenges, we introduce an end-to-end, LLM-driven pipeline that automatically constructs a structured and trustworthy social knowledge corpus $\mathcal{K}$.
This method ensures pedagogical traceability and supports diverse social skills at scale.



\subsubsection{\textbf{Theory-to-Practice Knowledge Organization}}
Effective social skills education requires more than isolated facts or generic tips; it demands a comprehensive framework that connects abstract principles, concrete examples, and hands-on practice. Most existing approaches treat these components in isolation, limiting transferability and depth of learning. We address this gap by designing a theory-to-practice knowledge framework across three pedagogical levels~\cite{education-esj-1998-shulman-theory-practice-education}. This method links what learners should do with why and how to do it in practice. This integration comprises three interrelated layers:
\begin{itemize}
    \item \textit{Strategic Theories} $\mathcal{K}_t$: Provides the conceptual "why" through abstract principles defining effective social interaction.
    \item \textit{Illustrative Case} $\mathcal{K}_c$: Bridges theory and practice by showing the "how" through concrete examples and case studies that demonstrate social principles in action.
    \item \textit{Practical Scenarios} $\mathcal{K}_s$: Enables the "doing" by immersing learners in hands-on simulations for active practice and assessment, thereby closing the "knowing-doing" gap.
\end{itemize}





\begin{figure}
    \centering
    \includegraphics[width=0.60\linewidth]{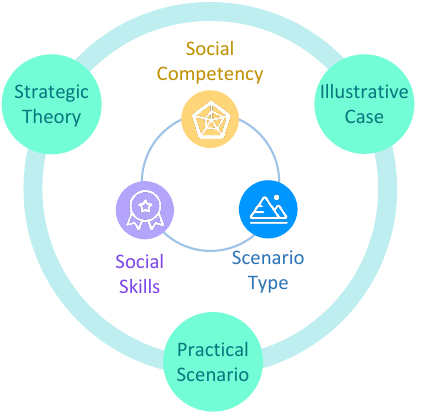}
    \vspace{-9pt}
    \caption{The proposed theory-to-practice social knowledge framework with multi-faceted categorization perspectives.}
    \vspace{-9pt}
    \label{fig:social-knowledge-corpus}
\end{figure}

\subsubsection{\textbf{Multi-faceted Semantic Categorization}}
Social interactions are highly diverse and context-dependent, requiring not only an understanding of specific skills but also their flexible application across various scenarios. Thus, we introduce a multi-faceted semantic categorization method to structure the knowledge corpus.
This method represents knowledge in a granular yet holistic way, facilitating flexible indexing and fine-grained retrieval of content based on each learner’s unique needs.
Specifically, each knowledge item is categorized along the following dimensions:
\begin{itemize}
    \item \textit{Social Competency} ($\mathcal{C}_c$): Foundational social-emotional abilities based on widely-used CASEL theory~\cite{sst-book-2025-sel-casel}, encompassing self-awareness, self-management, social awareness, relationship skills, and responsible decision-making.
    \item \textit{Social Skills} ($\mathcal{C}_s$): Actionable, specific skills such as conflict resolution, empathy, and negotiation. Drawing from the CASEL framework, we identify 34 essential social skills that support the five core competencies~\footnote{\url{https://casel.org/fundamentals-of-sel/what-is-the-casel-framework}}.
    \item \textit{Scenario Context} ($\mathcal{C}^t$): Contextual categories that are derived from social environments and relationship dynamics, such as workplace, romantic and public situations~\cite{sst-frontpsychol-2022-six-components-social-interaction}.
\end{itemize}
See Appendix~\ref{appendix:social-knowledge-categorization} for detailed descriptions of these categories.

\subsubsection{\textbf{LLM-Powered Corpus Construction Pipeline}}
To scalably extract pedagogical and traceable content of social knowledge, we introduce an automated pipeline leveraging LLM-powered multi-agent systems for efficient knowledge extraction and annotation~\cite{knowledge-www-2025-oneke}. This pipeline yields a structured and pedagogically grounded knowledge base, with the following principal key stages.




%

(a) \textit{Curated Data Collection.} To ensure both pedagogical value and rigor, we strategically aggregate high-quality, diverse materials on social skill. These sources include authoritative books, academic papers, news articles, and other reputable resources, thereby ensuring coverage of established theories and authentic scenarios.



(b) \textit{Schema-Guided Extraction}.
We first define the schema to represent three types of social knowledge items (see Appendix A.1 for detailed schema description). Guided by this schema, an extraction agent systematically transforms the unstructured collected data into a set of structured knowledge entities.

(c) \textit{Output Validation.} Then, we use a filtering agent to verify that the the extracted content adheres to the schema's format guidelines and the completeness across all required fields.

(d) \textit{Multi-faceted Tagging.}
To structure the knowledge semantically, a tagging agent annotates each knowledge item with multi-faceted categories, enabling nuanced organization.

(e) \textit{Knowledge Refinement.}
After tagging, a refinement agent reviews and revises both the content and metadata of each item to improve quality and accuracy. This step resolves inconsistencies, such as incomplete descriptions or incorrect tag assignments.

Through this multi-stage process, we construct a structured social knowledge corpus $\mathcal{K} = (\mathcal{K}^t, \mathcal{K}_c, \mathcal{K}_s)$, where each $k \in \mathcal{K}$ is associated with semantic labels $c_c \in \mathcal{C}_c$, $c_s \in \mathcal{C}_s$, and $c^t \in \mathcal{C}^t$.

\subsection{\textbf{User Social Profiling}}
Personalized social skills training requires a record of learner's goals and prior practice~\cite{profile-arxiv-2024-user-modeling-profiling}. We maintain a user record $U^t$ that separates self-reported information from observed interaction evidence.
\begin{itemize}
    \item \textit{Personal Information} $U_i$. Each user is initially characterized by optional self-reported background, preferences, and target social skills $S_U \subseteq S$. These attributes tailor scenario practices and guidance content.
    \item \textit{Practice History} $H_t$. The system records completed scenarios, retrieval traces, transcript evidence, reflection responses, and factual activity logs. These observations provide context for later scheduling and tutoring.
    \item \textit{Assessment Records} $D_t$. When a learner completes a real practice, SocialCoach stores evidence-linked strengths, weaknesses, and diagnostic notes. These records retain the supporting transcript spans and are not converted into an LLM-estimated proficiency transition.
\end{itemize}
After each completed session, the skill proficiency changes $\Delta_t$ result in state transitions, $U_p^{t+1} \leftarrow (U_p^{t}, \Delta_t)$.
During the learning journey, the practice scheduler conditions on the learner information $U_i$, stated goals $G$, transited state $U_p^t$, and observed practice history $H_t$. The resulting event $(a_t,p_t,\Delta_t,U_p^{t+1})$ is appended to $H_{t+1}$.




\subsection{Adaptive Practice Scheduling via Trajectory-Level Preference Optimization}
The development of social skills is experiential, and effective progress hinges on presenting each learner with the right practice at the right time.
While static curricula lack enough personalization, LLM-generated content risks pedagogical value.
Inspired by the judgment of human coaches, we introduce an adaptive practice scheduler and optimize complete practice pathways using agentic RL~\cite{wang2025-roll}. The scheduler operates over traceable corpus events and an explicitly simulated proficiency state. We separate the state-transition estimator from the pairwise pathway judge so that the same score is not used both to create state changes and to independently reward each session.

\subsubsection{\textbf{Prescription-Retrieval-then-Adaptation}.}
At each session $t$, we schedule a scenario practice $p^t$ with the following stages.

\textit{Practice Prescription}.
The prescription agent conditions on the learner's profile, stated goals, current simulated proficiency, and observed practice history to produce the next scenario requirement.
This requirement consists of a brief query, core skill constraints, and optional scenario attributes used as retrieval filters, such as situational context and relationship type.

\textit{Candidate Retrieval}.
The prescribed requirement serves as a query and filters to retrieve the most suitable scenario from the social knowledge corpus $K_s$. We combine tag-based filtering for pedagogical alignment with semantic search to rank eligible candidates. Optional constraints are relaxed in a fixed order when necessary; core skill constraints are retained, and an unsatisfied core constraint returns an explicit failure.

\textit{Context Adaptation}.
Finally, a context adaptation agent personalizes the retrieved scenario. Based on the learner’s profile $U^t$, the agent selects the most suitable role in the scenario for the learner's role-playing practice.
This step personalizes the retrieved context while remaining grounded in the structured and traceable corpus.

See Appendix \ref{appendix:prescription-schema} for the prescription schema and retrievability.

\subsubsection{\textbf{Cold-Start Pathway Environment}}

\begin{figure}[t]
    \centering
    \includegraphics[width=\linewidth]{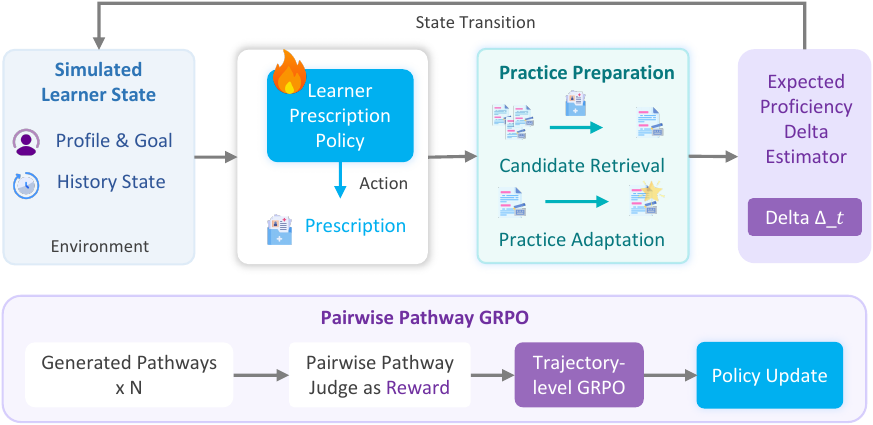}
    \vspace{-8pt}
    \caption{Prescription policy optimization with the pathway preference GRPO for cold-start personalization.}
    \vspace{-12pt}
    \label{fig:rl-training-retriever}
\end{figure}

Real longitudinal learning trajectories are scarce and costly to collect. We therefore build a cold-start environment from representative learner profiles and the frozen scenario corpus. It represents cognitive change only through a bounded expected transition conditioned on the selected practice; it does not generate learner utterances or treat the transition as observed human learning. The resulting state trajectory models short-horizon proficiency transitions within a scheduled sequence.

\textit{Representative Learner Profiles.}
We construct profiles from SocioVorse~\cite{dataset-arxiv-2025-socioverse} and PersonaHub~\cite{dataset-arxiv-2024-scaling-personas}. We embed demographic and personality descriptions with Qwen3-Embedding-8B~\cite{llm-arxiv-2025-qwen3-embedding}, use Maximal Marginal Relevance~\cite{profile-sigir-1998-mmr-diversity-reranking} to obtain a diverse subset, and apply the Hungarian algorithm~\cite{profile-nrlq-1955-hungarian-assignment-problem} for one-to-one matching. As in our original profile construction, each matched profile is assigned 3--5 target skills and starting proficiency scores $U_p^0$ on a $1$--$5$ scale. These scores instantiate heterogeneous cold-start states, which are synthetic assignments to simulate the initial states and goal.


\textit{Expected Proficiency Delta Estimator.}
To simulate the learning state transition, we employ an expected proficiency delta estimator $\text{LLM}_{E_\Delta}$ based on an LLM rubric judge to serve as a state transition proxy. Its rubric is pre-calibrated by experts to refine the criteria and anchor definitions, which considers profile and goal fit, target-skill coverage, fit between current proficiency and scenario difficulty, etc.
After retrieval, $\text{LLM}_{E_\Delta}$ receives the learner profile $U$, goals $G$, current target-skill state $U_p^t$, practice history $H_t$ and scheduled scenario practice $p_t$. It predicts the expected proficiency change assuming completing the practice,
\begin{equation}
\Delta_t=\text{LLM}_{E_\Delta}(U,G,U_p^t,H_t,p_t).
\end{equation}
Its output is constrained to a bounded nonnegative interval. Unsupported or weakly associated skills and severe difficulty mismatch receive zero delta; supported cross-skill changes have a lower cap than directly practiced skills. Let $\mathbf{m}(p_t)$ denote the mask of skills supported by the scenario. The simulated state transition is
\begin{equation}
U_p^{t+1}=\operatorname{clip}_{[1,5]}\!\left(U_p^t+\mathbf{m}(p_t)\odot\Delta_t\right),
\end{equation}
Here, $\odot$ denotes element-wise multiplication.
Appendix~\ref{appendix:human-evaluation-alignment} independently validates this proxy on held-out transition cases.


\textit{Dynamic Pathway State.}
At step $t$, the environment state is $s_t=(U,G,U_p^t,H_t)$, where $U$ and $G$ are learner information and goals, $U_p^t$ is the simulated proficiency vector, and $H_t$ is the append-only event history. The scenario corpus $\mathcal{K}_s$ and pathway horizon $T$ are fixed; prior coverage and remaining steps are derived deterministically from $H_t$ and $t$ rather than represented as separate policy inputs. The policy emits $a_t$, retrieval returns $p_t$, and the transition appends
$(a_t,p_t,\Delta_t, U_p^{t+1})$
to $H_t$. Thus, expected proficiency and observed practice history evolve across the pathway.


\subsubsection{\textbf{Pairwise Trajectory Reward and GRPO}}
In open-ended social learning, multiple pathways may produce cumulative and user-dependent outcomes, making statewise scoring and absolute trajectory rewards difficult to calibrate~\cite{zhang2026dynsessdynamicsessionlevelevaluation}. Therefore, we propose a pathway preference GRPO to train it with complete-path pairwise feedback, which provides a more reliable and discriminative optimization signal~\cite{zhang2026arenarlscalingrlopenended}.


For each initial profile, the policy samples a group of complete pathways $\{\tau_i\}_{i=1}^{M}$. For such an open-ended task, we use a Qwen3-235B judge $J_{\mathrm{train}}$ that compares pathway pairs using a fixed rubric for goal alignment, scenario fit, sequence coherence, appropriate coverage, and expected target-skill progression~\cite{jia2026openrubricsystemscaling}. Given the same learner profile and two pathway histories, its output is
\begin{equation}
y_{ij}=J_{\mathrm{train}}(U,G,\tau_i,\tau_j)\in\{1,0.5,0\},
\end{equation}
where the values denote win, tie, and loss for $\tau_i$. We randomize presentation order and compute each pathway's pairwise win rate $w_i=N_i^{-1}\sum_{j\in\mathcal{N}_i}y_{ij}$ over its compared opponents $\mathcal{N}_i$, with $N_i=|\mathcal{N}_i|$. We rank the $M$ pathways by $w_i$. Let $q_i\in[1,M]$ denote this rank from lowest to highest, with average ranks for ties. GRPO~\cite{rl-arxiv-2024-deepseekmath} uses the group reward and trajectory-level advantage
\begin{equation}
R_i=2\frac{q_i-1}{M-1}-1, \qquad
A_i=\frac{R_i-\mu_R}{\sigma_R+\epsilon}.
\end{equation}
The same $A_i$ is applied to the prescription tokens in $\tau_i$; no intermediate session receives a separate reward. Groups with no rank variation are skipped. We further fine-tune with multi-turn RL.

\paragraph{\textbf{Signal Separation}.}
The delta estimator only updates the simulated learning state. The pairwise judge supplies the trajectory preference used for optimization. Final pathway evaluation uses an independent pointwise $1$--$5$ judge that scores one generated pathway at a time, together with a blinded expert subset.

\vspace{-7pt}
\subsection{Pedagogical Coaching Scaffolding}
Effective social skill acquisition requires more than passive exposure to content. It demands an iterative loop of guided practice, diagnostic assessment, and personalized feedback grounded in pedagogical principles~\cite{sst-jritl-2019-social-emotional-learning-outlier}.
To address this, we introduce a three-stage coaching framework in SocialCoach that integrates immersive practice, evidence-linked practice assessment, and reflective tutoring.
By linking theory to practice, SocialCoach is designed to support deeper comprehension and application of the discussed strategies, helping address the "knowing–doing" gap.

\subsubsection{\textbf{Immersive Scenario-Based Practice}}
At each learning session $t$, for the scheduled scenario $k_s^t$, the social skill learning begins with immersive, goal-driven simulations designed to replicate real-world challenges. These tailored scenarios ensure engagement by creating a gamified practice environment.


\textit{Goal-Driven Scenario Simulation}.
The practice sessions are structured as goal-driven interaction games, where learners assume specific roles within challenging contexts~\cite{social-iclr-2024-sotopia}. Each session tasks the learner with achieving defined social objectives, such as mediating a conflict or negotiating a compromise, while interacting with LLM-driven agents representing other participants. These agents dynamically adapt their behavior and dialogue to the learner’s actions, ensuring a realistic and responsive simulation. Sessions conclude upon goal achievement or reaching the interaction limit, with conversation logs recorded as $L^t \leftarrow \text{simulation}(U^t, k_s^t)$.


\textit{Role-Playing Consistency Enhancement}
To maintain realism, we infer contextual features such as power dynamics, emotional tone, and implicit social norms of virtual agents based on extracted dialogues in the corpus. Then, we embed them into prompts to enhance the response consistency, aligning with their character roles and the scenario context, and enhancing the interaction immersion.

\subsubsection{\textbf{Practice Assessment}}
After each practice, SocialCoach identifies transcript-grounded strengths and weaknesses.

\textit{Social Behavior Diagnosis}.
Using an LLM-powered diagnostic agent, SocialCoach analyzes conversation logs $L^t$ to identify explicit strategies (e.g., active listening) and implicit reasoning (e.g., emotional awareness), i.e.,  $D^t_b = \text{LLM}_{\text{diagnosis}}(U^t, L^t)$. The agent detects and categorizes both positive and negative behaviors demonstrated during the simulation. These behaviors are mapped to CASEL competencies and targeted social skills, which enables precise identification of skill gaps and areas of improvement.

\textit{Deficit Attribution}.
Each weakness is classified as an \textit{acquisition deficit} (insufficient understanding) or a \textit{performance deficit} (difficulty applying known skills), i.e., $D^t_c= \text{LLM}_{\text{attribution}}(U^t,L^t,D^t_b)$~\cite{sst-guilford-2015-gresham-disruptive-behavior-disorders}. These are tutoring labels, not causal claims.

\subsubsection{\textbf{Knowledge-Grounded Reflective Tutoring}}
To close the coaching loop, SocialCoach delivers reflective, actionable feedback tailored to the learner’s specific needs. This stage aims to support learners' understanding and application of the discussed strategies.

\textit{Targeted Knowledge Retrieval}.
Based on the diagnostic assessment, SocialCoach retrieves relevant materials based on both tag-based filter and semantic embedding from the structured knowledge corpus, $\mathcal{K}^t = (\mathcal{K}^t_t, \mathcal{K}^t_c) \subset \mathcal{K}$. For identified \textit{performance deficits}, illustrative cases $\mathcal{K}^t_c \subseteq \mathcal{K}_c$ are provided to demonstrate skill application in analogous real-world contexts. Additionally, for \textit{acquisition deficits}, foundational theories $\mathcal{K}^t_t \subseteq \mathcal{K}^t$ are also retrieved to strengthen conceptual understanding.



\textit{Reflective Guidance Generation}.
Rather than just static feedback $D^t$ and $\mathcal{K}^t$, SocialCoach further adopts Socratic questioning~\cite{education-neurips-2024-socraticlm-personalized-teaching} to foster critical reflection, i.e., $G^t = \text{LLM}_{\text{guidance}}(U^t,L^t,D^t_b,D^t_c,\mathcal{K}^t)$. For example, it might ask: "Back to the situation of a case, what alternative strategies could you have employed to de-escalate the conflict?" This approach enables a reflective exercise, encouraging learners to analyze their actions and explore alternative strategies.

\vspace{-7pt}
\section{Experiments}
We conduct a comprehensive empirical study of SocialCoach to assess the quality of its knowledge corpus, the judge-rated quality of its adaptive scenario scheduling, and the perceived value of its pedagogically grounded coaching. Generated outputs, prompt templates, and evaluation materials are available in the \href{https://github.com/GeminiLight/social-coach-artifact}{public SocialCoach materials repository}.

\subsection{System Implementation}

\subsubsection{\textbf{Data Collection Method}}

For the research goal in this study, we construct the social knowledge corpus by curating books from recognized lists and category rankings. First, we identify relevant books using category rankings from platforms like Amazon Books' self-help section\footnote{\href{https://www.amazon.com/amz-books/store}{https://www.amazon.com/amz-books/store}} (e.g., relationships, emotional intelligence) and BookAuthority.org\footnote{\href{https://bookauthority.org/categories}{https://bookauthority.org/categories}} (e.g., leadership, empathy, social intelligence).
Then, we gather the available files of these books from the Open Library website\footnote{\href{https://openlibrary.org}{https://openlibrary.org}}.
After a manual review for relevance and quality, we selected 200 books to ensure a focused and comprehensive corpus.
The final corpus includes many widely cited works such as ``Nonviolent Communication: A Language of Life" and ``Emotional Intelligence: Why It Can Matter More Than IQ".

\subsubsection{\textbf{System Implementation Details}}
We support various LLMs as agent backends, including GPT-5, Gemini-3-Pro, DeepSeek-R1-0528 (DSR1), Qwen3-235B, and Qwen3-8B. Concretely, for the corpus construction pipeline, we employ Qwen3-235B as the backend, Chroma as the vector database, and Qwen3-Embedding-8B~\cite{llm-arxiv-2025-qwen3-embedding} as the embedding model for candidate retrieval. Regarding the representative profile used in simulated environments, we visualize the distribution to show their consistency and diversity. See Appendix~\ref{appendix:profile-distribution-visualization} for details.
Next, for the prescription agent, we adopt Qwen3-8B as the base LLM and optimize it with pathway preference GRPO~\cite{rl-arxiv-2024-deepseekmath}, implemented using the ROLL framework~\cite{wang2025-roll} for multi-turn agentic reinforcement learning. In a simulated environment, both the proficiency estimator $\text{LLM}_{E_\Delta}$ and pairwise reward judge use Qwen3-235B. See more training details in Appendix~\ref{appendix:training-setting-details}.
Finally, for coaching scaffolding, this service layer supports multiple LLMs and uses GPT-5 by default.


\subsubsection{\textbf{Human Expert Panel}}
All expert-based evaluations use the same panel of three workplace practitioners, each with 5--8 years of relevant professional experience in social-skills development and emotional-intelligence-related practice. For each task, they review the task-specific rubric and anchors before independently evaluating the items; model identities, automatic scores, and experimental conditions are hidden whenever applicable.






\begin{table}[t!]
\centering
\caption{Corpus Statistics and Tag Diversity.}
\vspace{-7pt}
\resizebox{.5\textwidth}{!}{
\begin{tabular}{cccccccc}
\toprule
\multirow{2}{*}{\textbf{Type}}
& \multicolumn{2}{c}{\textbf{Overall Statistics}}
& \multicolumn{3}{c}{\textbf{Norm. Shannon Entropy}} \\
\cmidrule(lr){2-3} \cmidrule(lr){4-6}
 & \textbf{Size} & \textbf{Avg. Words} & \textbf{Competence} & \textbf{Skill} & \textbf{Context} \\
\midrule
Theory  & 14,186 & 150.21 & 0.81 & 0.74 & N/A \\
Case & 23,521 & 86.92 & 0.84 & 0.77 & 0.79 \\
Scenario & 5,463 & 372.45 & 0.86 & 0.89 & 0.82 \\
\bottomrule
\end{tabular}
}
\vspace{-7pt}
\label{tab:corpus_statistics}
\end{table}

\begin{table}[t!]
\centering
\caption{Mean Expert Likert Ratings on Corpus.}
\vspace{-7pt}
\resizebox{.5\textwidth}{!}{
\begin{tabular}{ccccccc}
\toprule
\multirow{2}{*}{\textbf{Type}}
& \multicolumn{3}{c}{\textbf{Mean Tag Ratings}}
& \multicolumn{3}{c}{\textbf{Content Pedagogical Value}} \\
\cmidrule(lr){2-4} \cmidrule(lr){5-7}
 & \textbf{Competence} & \textbf{Skill} & \textbf{Context}
 & \textbf{Soundness} & \textbf{Accessibility} & \textbf{Pedagogy}
\\
\midrule
Theory  & 4.78 & 4.62 & - & 4.08 & 3.81 & 3.26 \\
Case & 4.68 & 4.55 & 3.86 & 4.06 & 3.98 & 3.40 \\
Scenario & 4.92 & 4.82 & 4.63 & 4.83 & 4.41 & 4.44 \\
\bottomrule
\end{tabular}
}
\vspace{-7pt}
\label{tab:corpus_quality_expert_ratings}
\end{table}

\subsection{\textbf{Analysis of Knowledge Corpus}}
Our automated pipeline produces a corpus of 43,170 structured knowledge entries, each tagged with multifaceted semantics.
To assess the quality, diversity, and pedagogical value of this corpus, we conducted the following multi-faceted analysis.

\subsubsection{\textbf{Overall Corpus Statistics}}
As shown in Table~\ref{tab:corpus_statistics}, this corpus is composed of 14,186 theories, 23,521 cases, and 5,463 scenarios. We analyze the textual depth of each category by calculating the average word count, revealing that scenarios are considerably longer than theories and cases. This difference reflects their pedagogical roles: scenarios provide detailed immersive practice, while theories and cases offer concise conceptual knowledge and examples.

\subsubsection{\textbf{Tag Diversity \& Distribution}}
To estimate whether the corpus is well-balanced, we evaluated the diversity of our semantic tags using normalized Shannon entropy, which measures the uniformity of a distribution on a $[0,1]$ scale. As shown in Table~\ref{tab:corpus_statistics}, the entropy values indicate that the corpus covers a broad range of social situations rather than being limited to a few contexts.
A detailed breakdown of the tag distributions is as follows:

\subsubsection{\textbf{Expert Likert Ratings}}
To assess tag quality and pedagogical value, the expert panel independently rated a stratified random sample of 150 corpus entries (50 per knowledge type). Items were presented in randomized order, using predefined 1--5 rating criteria. The manual validation yielded an average Fleiss' Kappa of 0.63 across assessments. See Appendix~\ref{appendix:expert-likert} for details.

\subsection{Evaluation on System Quality}
We evaluate on 200 profiles held out from policy training. Each profile combines demographic context, personality description, stated target skills, and a synthetic initial proficiency state. To ensure a matched comparison, all scheduling methods receive the same initial states, and all transition-aware configurations use the same frozen delta estimator. Its cumulative delta is reported only as a planning diagnostic; the primary final outcomes remain pathway quality and expert alignment.
Following existing studies~\cite{llm-arxiv-2024-survey-llm-as-a-judge,education-www-2025-gen-mentor,simulation-arxiv-2025-sotopia-rl}, we use GPT-5~\cite{openai2025gpt5systemcard} as an independent pointwise evaluator. It sees one pathway at a time and assigns anchored $1$--$5$ ratings.


\subsubsection{\textbf{Evaluation on Practice Scheduling}}
We compare the following baselines: (a) \textit{Profile Retrieval}~\cite{retrieval-emnlp-2023-lapdog} directly ranks scenarios against the serialized profile and goals; (b) \textit{P-RAG$_{\text{model}}$}~\cite{retrieval-aaai-2026-personalize-before-retrieve} generates a fixed pathway from a personalized query; (c) \textit{HA-RAG}~\cite{retrieval-findings-2024-haconvdr} sequentially schedules practices conditioned each query on persistent practice history.
All methods use the same initial profile, corpus snapshot, prescription schema, retrieval interface, and 10-session horizon. We instantiate P-RAG and HA-RAG with Qwen3-8B, Qwen3-235B, DeepSeek-R1-0528, and GPT-5.

\paragraph{\textbf{Baseline Comparison.}}
Results are shown in Table~\ref{tab:evaluation-practice-pathway}. Across the four pathway-quality dimensions, SocialCoach outperforms these baselines, even based on a better backbone model.
Paired bootstrap analysis shows that SocialCoach improves over HA-RAG$_{\text{GPT-5}}$ by 0.47 points on average across the four pathway-quality dimensions (95\% CI [0.40, 0.54], $p<0.01$).
With Qwen3-8B as the shared backbone, SocialCoach also reduces the first-attempt retrieval failure rate from 3.5\% to 0.0\%, indicating better alignment between its prescriptions and the available corpus.


\paragraph{\textbf{Ablation Study.}}
We further evaluate three match ablations:
(a) \emph{w/o Proficiency Transition} fixes $U_p^t=U_p^0$ while retaining the growing practice history, isolating the contribution of progression-aware proficiency updates. (b) \emph{w/o Pairwise Reward} replaces A/B/Tie pathway comparisons with independently assigned $1$--$5$ pathway scores while preserving trajectory-level credit assignment, isolating the effect of feedback form. (c) \emph{w/o GRPO Training} retains the full progression-aware state and frozen $E_\Delta$ but uses the untuned Qwen3-8B policy without judge feedback or policy updates, isolating the effect of trajectory-level policy optimization. All other components and computational budgets are held fixed unless directly ablated.
All three ablations underperform the full model across the pathway-quality dimensions. Removing GRPO training causes the largest degradation, while freezing proficiency transitions also consistently reduces performance. The pointwise-reward variant performs worse than its pairwise counterpart, supporting the benefit of pairwise feedback beyond trajectory-level credit alone.

\paragraph{\textbf{Cross-Judge Robustness.}}
We further pre-specify a cross-family robustness check using Qwen3-235B and DeepSeek-R1-0528 in Appendix~\ref{appendix:cross-judge-robustness}, which shows the evaluation robustness.

\begin{table}[t]
\centering
\caption{Evaluation results on practice scheduling. Entries are mean $\pm$ SD across three independent runs.}

\vspace{-8pt}
\label{tab:evaluation-practice-pathway}
\resizebox{\columnwidth}{!}{%
\begin{tabular}{lcccc}
\toprule
\textbf{Method} & \textbf{Goal Fit} & \textbf{Pers.} & \textbf{Ped. Value} & \textbf{Coher.} \\
\midrule
Profile Retrieval & 2.92 {\scriptsize$\pm$ 0.16} & 3.04 {\scriptsize$\pm$ 0.13} & 2.78 {\scriptsize$\pm$ 0.14} & 3.12 {\scriptsize$\pm$ 0.17} \\
P-RAG$_{\text{Qwen3-8B}}$ & 3.24 {\scriptsize$\pm$ 0.06} & 3.15 {\scriptsize$\pm$ 0.08} & 3.12 {\scriptsize$\pm$ 0.16} & 3.40 {\scriptsize$\pm$ 0.17} \\
P-RAG$_{\text{Qwen3-235B}}$ & 3.36 {\scriptsize$\pm$ 0.16} & 3.46 {\scriptsize$\pm$ 0.09} & 3.30 {\scriptsize$\pm$ 0.05} & 3.56 {\scriptsize$\pm$ 0.16} \\
P-RAG$_{\text{DeepSeek-R1-0528}}$ & 3.45 {\scriptsize$\pm$ 0.14} & 3.34 {\scriptsize$\pm$ 0.16} & 3.32 {\scriptsize$\pm$ 0.18} & 3.58 {\scriptsize$\pm$ 0.06} \\
P-RAG$_{\text{GPT-5}}$ & 3.48 {\scriptsize$\pm$ 0.10} & 3.40 {\scriptsize$\pm$ 0.17} & 3.38 {\scriptsize$\pm$ 0.05} & 3.64 {\scriptsize$\pm$ 0.06} \\
HA-RAG$_{\text{Qwen3-8B}}$ & 3.46 {\scriptsize$\pm$ 0.17} & 3.38 {\scriptsize$\pm$ 0.15} & 3.34 {\scriptsize$\pm$ 0.12} & 3.61 {\scriptsize$\pm$ 0.07} \\
HA-RAG$_{\text{Qwen3-235B}}$ & 3.60 {\scriptsize$\pm$ 0.04} & 3.72 {\scriptsize$\pm$ 0.17} & 3.55 {\scriptsize$\pm$ 0.07} & 3.82 {\scriptsize$\pm$ 0.17} \\
HA-RAG$_{\text{DeepSeek-R1-0528}}$ & 3.74 {\scriptsize$\pm$ 0.11} & 3.64 {\scriptsize$\pm$ 0.07} & 3.60 {\scriptsize$\pm$ 0.09} & 3.86 {\scriptsize$\pm$ 0.15} \\
HA-RAG$_{\text{GPT-5}}$ & 3.72 {\scriptsize$\pm$ 0.05} & 3.70 {\scriptsize$\pm$ 0.06} & 3.66 {\scriptsize$\pm$ 0.05} & 3.90 {\scriptsize$\pm$ 0.09} \\
\midrule
\textbf{SocialCoach$_{\text{Qwen3-8B}}$} & \textbf{4.22 {\scriptsize$\pm$ 0.13}} & \textbf{4.18 {\scriptsize$\pm$ 0.14}} & \textbf{4.15 {\scriptsize$\pm$ 0.15}} & \textbf{4.32 {\scriptsize$\pm$ 0.06}} \\
\midrule
w/o GRPO Training & 3.72 {\scriptsize$\pm$ 0.10} & 3.61 {\scriptsize$\pm$ 0.16} & 3.58 {\scriptsize$\pm$ 0.06} & 3.75 {\scriptsize$\pm$ 0.16} \\
w/o Proficiency Transition & 4.04 {\scriptsize$\pm$ 0.12} & 3.96 {\scriptsize$\pm$ 0.09} & 3.92 {\scriptsize$\pm$ 0.15} & 4.14 {\scriptsize$\pm$ 0.07} \\
w/o Pairwise Reward & 4.02 {\scriptsize$\pm$ 0.14} & 4.06 {\scriptsize$\pm$ 0.05} & 3.91 {\scriptsize$\pm$ 0.06} & 4.12 {\scriptsize$\pm$ 0.17} \\
\bottomrule
\end{tabular}%
}
\vspace{-10pt}
\end{table}

\begin{figure*}[t]
  \centering
\includegraphics[width=0.90\linewidth]{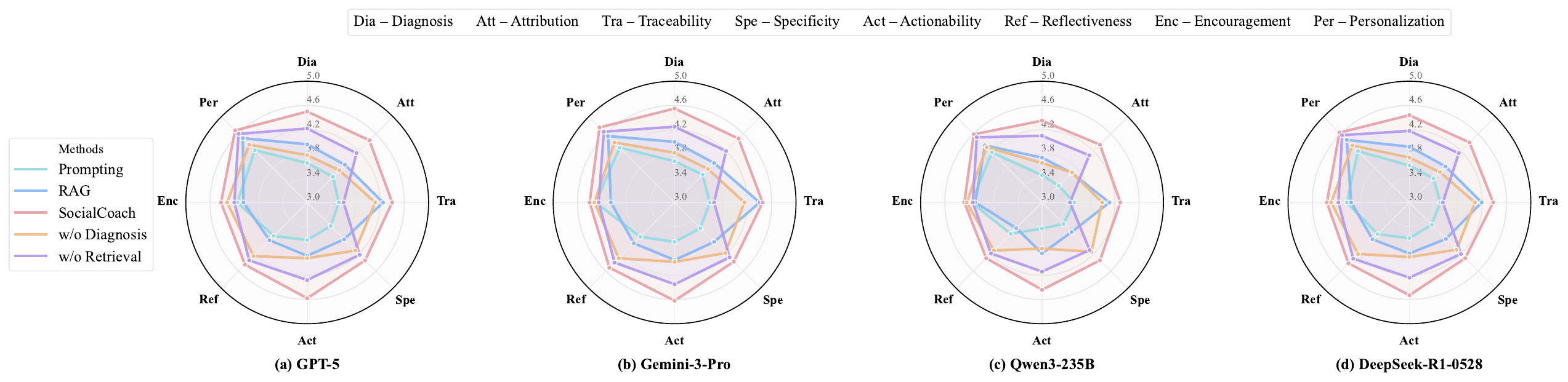}
    \vspace{-10pt}
 \caption{Evaluation results on tutoring guidance.}
    \vspace{-10pt}
 \label{fig:results-pedagogical-tutoring}
\end{figure*}

\subsubsection{\textbf{Evaluation on Pedagogical Tutoring}}

To evaluate the pedagogical feedback, we use GPT-5~\cite{openai2025gpt5systemcard} to generate synthetic interaction traces for 200 practice sessions. These traces are evaluation inputs rather than observations of learner behavior or learning change. We compare SocialCoach against four baselines: (a) Prompting LLM that provides feedback directly, (b) RAG model that retrieves from our corpus, and two ablated variants of our system: (c) one without assessment workflow, and (d) another without retrieval module.
Safety screening excludes high-risk and crisis-oriented scenarios, restricting SocialCoach to low-stakes practice.
We evaluate guidance along eight dimensions: Diagnosis (problem identification), Attribution (deficit-classification quality), Traceability (source faithfulness), Specificity (relevance to user actions), Actionability (clarity of improvements), Reflectiveness (Socratic questioning), Encouragement (supportive tone), and Personalization (fit to the learner profile).

As shown in Figure~\ref{fig:results-pedagogical-tutoring}, SocialCoach receives higher scores than the baselines and ablated models across most metrics. Its higher Diagnosis and Attribution ratings support separating behavior identification from provisional deficit classification.
Additionally, we observe that both retrieval-based methods, such as SocialCoach and RAG, achieve high scores in Traceability. This suggests that anchoring feedback in a verifiable knowledge corpus can improve perceived reliability.
Furthermore, SocialCoach also receives higher scores on more integrative metrics, such as Actionability and Personalization. This supports the value of combining component design with educational domain knowledge.
Overall, these results suggest that SocialCoach can deliver personalized, pedagogical guidance within a comprehensive coaching loop under our synthetic-input and judge-rated evaluation setting.

\subsubsection{\textbf{Alignment with Human Evaluation}}
We further conduct several blinded expert studies to validate the alignment of automated evaluation results with human judgment. (a) \textit{Proficiency-Transition Estimator.} On 100 held-out transition cases, the estimator shows moderate agreement with expert labels (weighted $\kappa=0.56$). (b) \textit{Pairwise Training Judge.} On 100 held-out pathway pairs, the judge reaches 74\% exact A/B/Tie agreement with expert-majority judgments ($\kappa=0.51$). (c) \textit{Expert Pathway Preference.} On a separate set of 100 held-out profiles, experts favor SocialCoach$_{\text{Qwen3-8B}}$ over HA-RAG$_{\text{GPT-5}}$, with a tie-adjusted preference rate of 68\% (95\% CI [62\%, 74\%]). (d) \textit{Pointwise Evaluator.} When scoring the same anonymous pathway and tutoring outputs, the evaluator correlates moderately with expert ratings across all four pathway-quality dimensions ($\rho\in[0.55,0.64]$, all $p<0.01$) and six of eight tutoring criteria; Specificity and Encouragement do not reach statistical significance.
Together, these results support the automatic signals as planning, training, and evaluation proxies, but do not establish human learning gains. See Appendix~\ref{appendix:human-evaluation-alignment} for their details.

\subsubsection{\textbf{Case Study of Learning Journey}} In addition to our quantitative analysis and LLM-based evaluations, we further provide case studies in Appendix~\ref{appendix:case-studies}, which trace a concrete learning journey to illustrate system behavior and perceived pedagogical value.



\vspace{-6pt}
\subsection{End-to-end User Study}
\subsubsection{Practical Platform Deployment}
We instantiated SocialCoach in an internally deployed research platform.
The platform features a mobile-first React frontend and a Python backend. SocialCoach serves as its core engine via algorithmic APIs that power adaptive practice scheduling, the LLM-driven simulation arena, and the coaching loop.
We provide its key interfaces in Appendix~\ref{appendix:product-demo}.
The platform was made available to an internal beta cohort before the user study, offering a gamified environment for social skill learning.



\subsubsection{{User Study Analysis}}
With institutional review board approval, we conducted a 10-practice study with 50 participants (25 university students, 25 early-career professionals; aged 20–30).
Participants were instructed to use the platform for at least 10 practices.
At the end of the study period, we administered a post-study perceived-usefulness questionnaire and conducted semi-structured interviews. See Appendix~\ref{appendix:user-study} for the results.

Additionally, we also conducted a randomized pilot ($n=10$) over that yielded numerically higher adaptive-group means for engagement, expert-rated pedagogical fit, and immediate and delayed strategy recall. See Appendix~\ref{appendix:learning-experience-study} for details.

\vspace{-7pt}
\section{Conclusion}
In this paper, we introduced SocialCoach, an LLM-powered agentic tutoring and practice framework for personalized social skill learning. We formulate its central personalization problem as cold-start, retrieval-constrained sequential practice scheduling. Our approach couples retrieval-grounded prescriptions, a bounded Expected Proficiency Delta Estimator that updates the synthetic target-skill state, and pairwise trajectory-level GRPO over complete pathways. The full system instantiates this scheduling approach with a theory-to-practice corpus and a downstream practice and reflective tutoring loop. Our evaluations demonstrate the quality and diversity of the constructed corpus and show that SocialCoach outperforms retrieval-based baselines and matched ablations in pathway scheduling and pedagogical tutoring. Blinded expert studies further support the reliability of its planning, training, and evaluation proxies and favor its generated pathways over the strongest baseline. Platform deployment and user studies provide early evidence of engagement, perceived usefulness, and improved learning experience. Together, these findings position SocialCoach as a practical and scalable framework for personalized social skill learning.


\bibliographystyle{ACM-Reference-Format}
\bibliography{main}

\appendix
\section{Details of Framework Design}






\subsection{Knowledge Categorization Details}
\label{appendix:social-knowledge-categorization}

To ensure our knowledge corpus is pedagogically sound and easily retrievable, we structure it using a multifaceted categorization method. Each piece of knowledge is tagged along three key dimensions: its associated CASEL competency, the specific social skill it addresses, and the scenario type in which it applies.

\subsubsection{\textbf{Associated Social Competencies}}
In this work, we adopt the Collaborative for Academic, Social, and Emotional Learning (CASEL) framework~\cite{sst-book-2025-sel-casel}, which is recognized as one of the most popular SEL frameworks by the Explore SEL project\footnote{\href{http://exploresel.gse.harvard.edu/com/domains/}{http://exploresel.gse.harvard.edu/compare-domains/}} of Harvard Graduate School of Education.
CASEL provides a widely acknowledged foundation for understanding social-emotional learning and identifies five core competencies: Self-Awareness, Self-Management, Social Awareness, Relationship Skills, and Responsible Decision-Making. These competencies represent the broad, overarching abilities crucial for navigating social interactions effectively.

\subsubsection{\textbf{Associated Social Skills}}
To bridge the gap between CASEL's abstract competencies and practical application, the CASEL office further categorizes knowledge by associated Social Skills. These are defined as granular, observable, and teachable behaviors—such as specific communication techniques or actions—that individuals can practice to demonstrate proficiency within the broader CASEL domains. Our system tags content with these fine-grained skills to enable the precise targeting of learning objectives and assessment criteria. The relationship between CASEL competencies and their corresponding social skills is detailed in Table \ref{tab:casel_competencies_and_skills}.

\subsubsection{\textbf{Scenario Contextual Types}}
Since the usefulness of a social skill often depends on its context, we also categorize scenarios by Scenario Type. This categorization is hierarchical to provide nuanced classification. First, we assign a high-level category describing the general environment where the interaction occurs (e.g., Workplace, Family). Next, we refine this with a more detailed sub-category (e.g., Office, Parent-child). This hierarchical approach, detailed in Table~\ref{tab:contextual_mapping}, helps learners practice skills in contexts directly relevant to their personal and professional lives.



\begin{table}[h!]
\centering
\caption{The list of social skills grouped by their main related competencies under the CASEL framework.}
\label{tab:casel_competencies_and_skills}
\renewcommand{\arraystretch}{1.3}
\resizebox{0.45\textwidth}{!}{
\begin{tabular}{p{2.5cm}|p{\dimexpr\columnwidth-2.5cm-2\tabcolsep-2\arrayrulewidth\relax}}
\toprule
\textbf{Competencies} & \textbf{Social Skills} \\
\midrule
\multirow{7}{=}{Self-Awareness} & Identifying emotions \\
& Social and cultural identity \\
& Recognizing strengths \\
& Growth mindset \\
& Self-efficacy \\
& Examining bias \\
& Sense of purpose \\
\midrule
\multirow{8}{=}{Self-Management} & Emotion regulation \\
& Impulse control \\
& Stress management \\
& Self-discipline and motivation \\
& Perseverance \\
& Goal-setting \\
& Organizational skills \\
& Initiative and Agency \\
\midrule
\multirow{6}{=}{Social Awareness} & Perspective-taking \\
& Empathy and compassion \\
& Expressing gratitude \\
& Appreciating diversity \\
& Identifying social norms and demands \\
& Sense of belonging \\
\hline
\multirow{8}{=}{Relationship Skills} & Communication \\
& Cultural competence \\
& Building relationships \\
& Teamwork and working cooperatively \\
& Resolving conflicts \\
& Helping/Seeking help \\
& Leadership \\
& Standing up for the rights of others \\
\midrule
\multirow{5}{=}{Responsible Decision-Making} & Demonstrating curiosity and open-mindedness \\
& Identifying and solving problems \\
& Analyzing situations and consequences \\
& Ethical responsibility \\
& Reflecting on one's role to promote individual and collective well-being \\
\bottomrule
\end{tabular}
}
\end{table}

\begin{table}[h!]
\centering
\caption{7 scenario contextual categories detailed in 26 types.}
  \vspace{-10pt}
\label{tab:contextual_mapping}
\resizebox{.5\textwidth}{!}{
\begin{tabular}{ll}
\toprule
\textbf{Contextual Category} & \textbf{Detailed Contextual Types} \\
\midrule
Workplace & Office, meetings, job interviews, professional collaboration \\
Family & Parent-child, siblings, home interactions, Spouse \\
Friendship & Informal, emotional bonding, casual talk \\
Romantic & Dating, partner conflicts, affection sharing \\
Education & Classroom, peer discussion, teacher-student roles \\
Public/Stranger & Store, transport, unknown people, Park, Courtroom \\
Party/Social & Events, birthdays, wine parties, social mingling \\
\bottomrule
\end{tabular}
}
  \vspace{-10pt}
\end{table}

\subsection{Prescription Schema and Retrievability}
\label{appendix:prescription-schema}

We provide a formal description of the prescription action schema and the retrievability mechanism for scenario practices.

\definecolor{eclipseStrings}{RGB}{42,0,255}
\definecolor{eclipseKeywords}{RGB}{127,0,85}
\colorlet{numb}{magenta!60!black}

\lstdefinelanguage{json}{
    basicstyle=\small\ttfamily,
    commentstyle=\color{eclipseStrings},
    stringstyle=\color{eclipseStrings},
    identifierstyle=\color{eclipseKeywords},
    numbers=left,
    numberstyle=\tiny,
    stepnumber=1,
    numbersep=8pt,
    showstringspaces=false,
    breaklines=true,
    frame=lines,
    backgroundcolor=\color{gray!5},
    string=[s]{"}{"},
    comment=[l]{:\ "},
    morecomment=[l]{:"},
    literate=
     *{0}{{{\color{numb}0}}}{1}
      {1}{{{\color{numb}1}}}{1}
      {2}{{{\color{numb}2}}}{1}
      {3}{{{\color{numb}3}}}{1}
      {4}{{{\color{numb}4}}}{1}
      {5}{{{\color{numb}5}}}{1}
      {6}{{{\color{numb}6}}}{1}
      {7}{{{\color{numb}7}}}{1}
      {8}{{{\color{numb}8}}}{1}
      {9}{{{\color{numb}9}}}{1}
}

\begin{figure}[t]
\centering
\begin{lstlisting}[
    language=json,
    caption={An example of a structured prescription.},
    label={lst:prescription_agent},
    basicstyle=\footnotesize\ttfamily,
    frame=tb,
    aboveskip=3mm,
    belowskip=3mm,
    showstringspaces=false,
    keywordstyle=\color{blue},
    stringstyle=\color{teal},
    breaklines=true
]
{
  "query": "Practice a respectful disagreement during a project meeting.",
  "core_constraints": {
    "target_skills": ["Conflict Resolution"],
    "contexts": ["Workplace"]
  },
  "optional_constraints": {
    "related_skills": ["Teamwork"],
    "relationship_types": ["Senior", "Peer"]
  }
}
\end{lstlisting}
\end{figure}

\subsubsection{Action Schema Components}
The action $a^{t}$ is a structured scenario requirement with a short retrieval query, core constraints, and optional preferences. Core constraints define what the selected practice must satisfy; optional fields refine context or relationship type when matching scenarios exist. All labels come from the frozen corpus taxonomy, which keeps retrieval behavior auditable and limits unconstrained query programs.

\subsubsection{Retrievability Guarantee}
Strict filtering can produce an empty candidate set. We therefore distinguish core constraints, such as the target skill, from optional context preferences. Retrieval first enforces both sets. If no candidate is found, optional constraints are relaxed in a fixed order and every change is recorded. Core constraints are never removed; if they cannot be satisfied, the system returns an explicit failure rather than presenting an unrelated scenario as suitable.

\section{Details on Experimental Setting}

\subsection{Descriptions of Training Settings}
\label{appendix:training-setting-details}
We train the prescription policy for 200 steps with a KL-loss coefficient of 0.001. For each prescription response, it has a maximum generation length of 2048 tokens. We configure the parallel environment manager with a batch size of 64 and a group size of 8.
We construct 1,000 representative training profiles with 3--5 target skills and synthetic starting proficiency scores on a $1$--$5$ scale, and optimize 10-session pathways by sampling eight pathways per GRPO group. After each retrieval, a rubric-based expected proficiency delta estimator with the frozen Qwen3-235B updates only the skills supported by the selected scenario, which assumes practice completion and does not generate a learner response. A frozen Qwen3-235B model as judge then provides A/B/Tie comparisons between complete pathways using both traceable scenario evidence and the resulting state trajectory. These training experiments are conducted on a cluster of 16 NVIDIA H20 GPUs with a learning rate of $1.0 \times 10^{-6}$.

\subsection{\textbf{Visualization on Profile Distribution}}
\label{appendix:profile-distribution-visualization}
\begin{figure}
    \centering
    \includegraphics[width=0.80\linewidth]{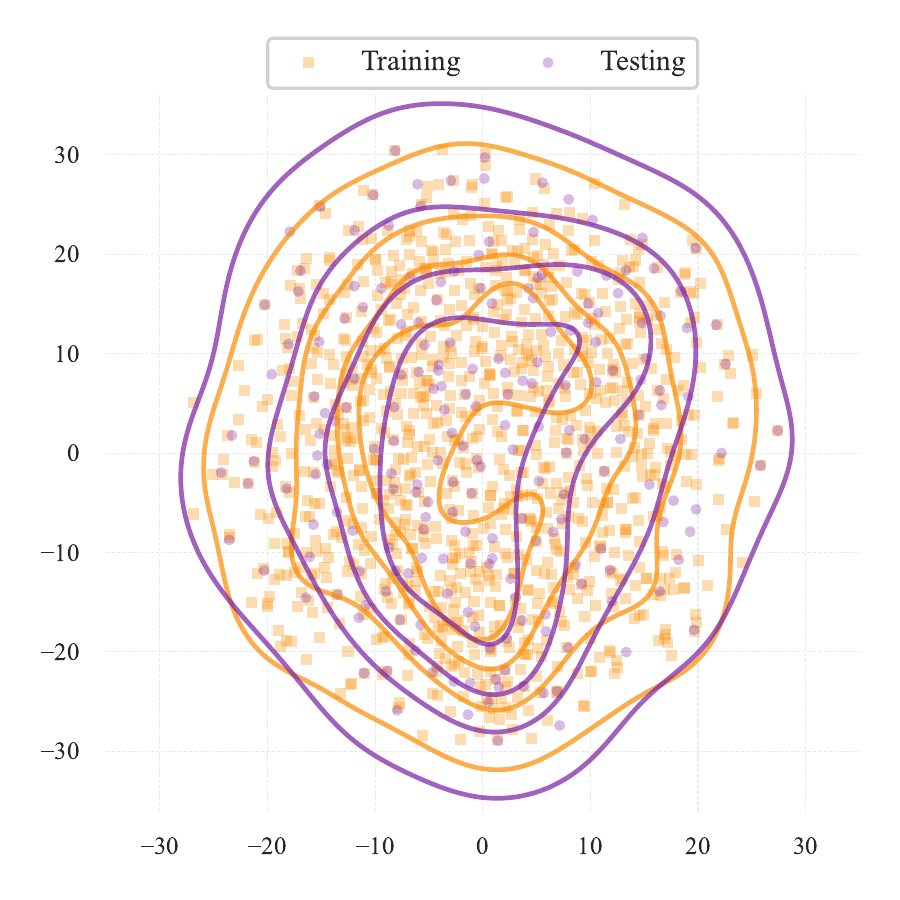}
    \vspace{-10pt}
    \caption{Representative learner profile distribution.}
  \vspace{-10pt}
    \label{fig:representative-learner-profile-distribution}
\end{figure}

We visualize the distribution of representative learner profile embeddings using t-SNE, as shown in Figure~\ref{fig:representative-learner-profile-distribution}.

\section{Analysis on Knowledge Corpus}

\subsection{Visualization on Corpus Distribution}
\begin{figure}
    \centering
    \includegraphics[width=0.96\linewidth]{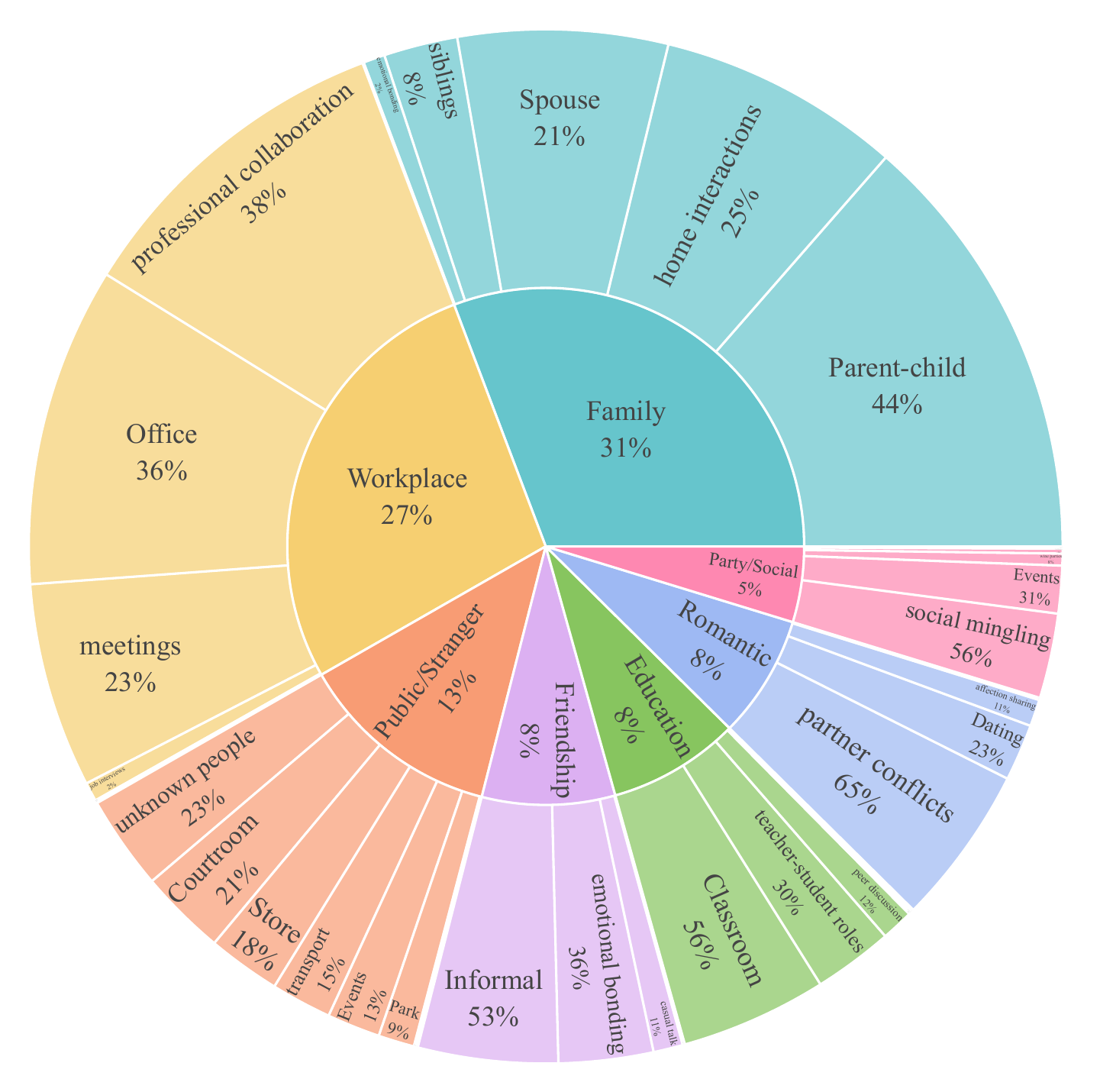}
    \vspace{-10pt}
    \caption{Distribution of scenario contextual types.}
  \vspace{-10pt}
    \label{fig:corpus-scenario-context-types}
\end{figure}

\begin{figure}
    \centering
    \includegraphics[width=0.90\linewidth]{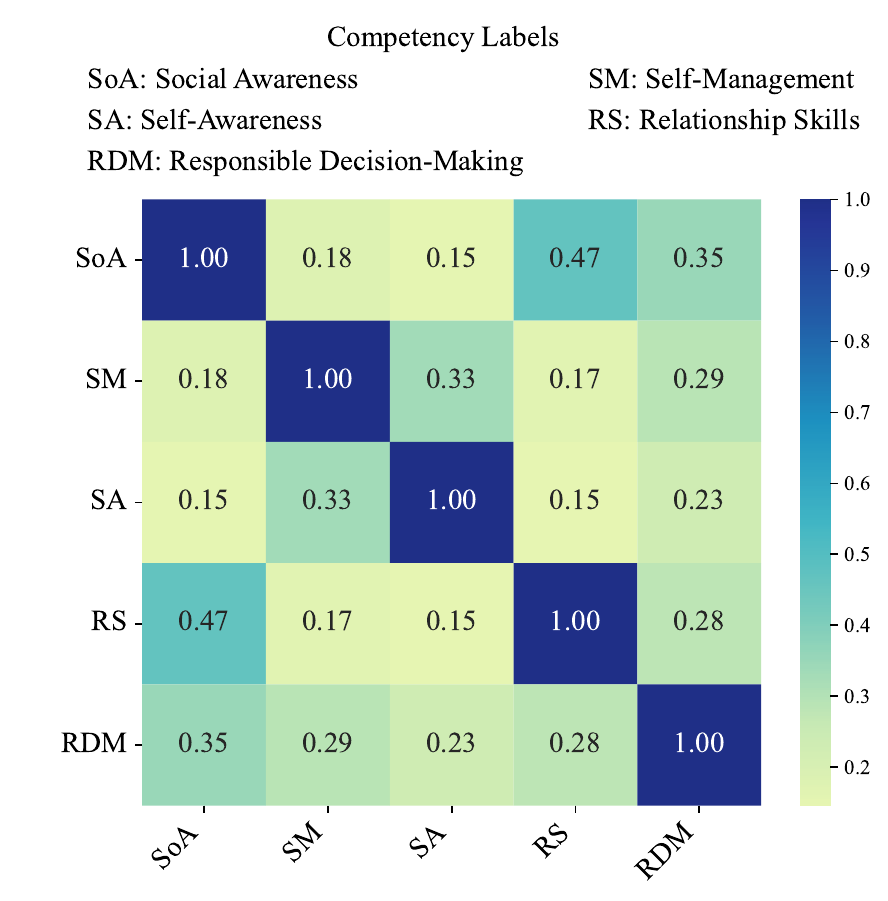}
    \vspace{-10pt}
    \caption{Co-occurrence heatmap of social competencies}
    \label{fig:corpus-competency-cooccurrence-heatmap}
\end{figure}

\begin{figure}
    \centering
    \includegraphics[width=0.96\linewidth]{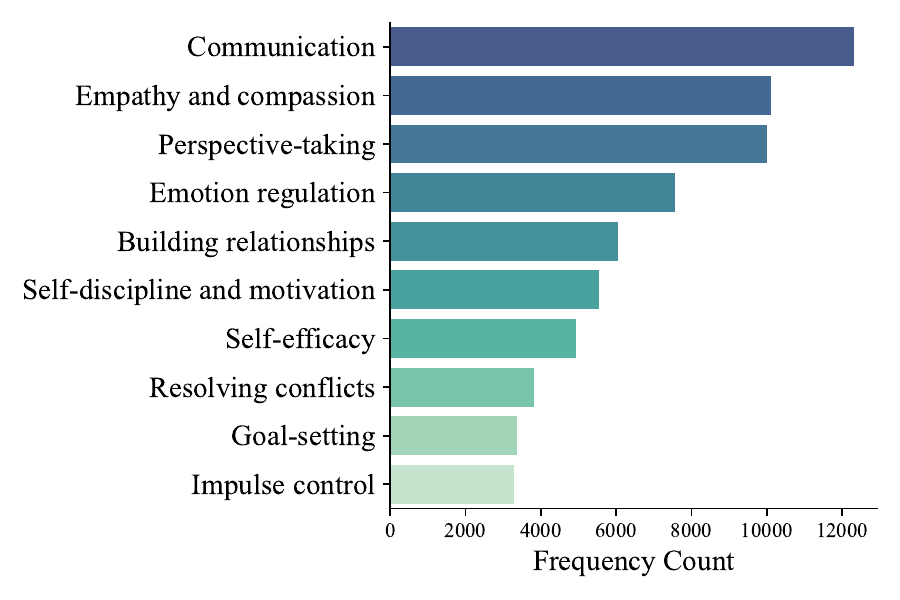}
    \vspace{-10pt}
    \caption{Top 10 social skills by frequency.}
    \label{fig:corpus-top-10-social-skills}
\end{figure}

\textit{Social Competency Co-occurrence}.
The heatmap in Figure~\ref{fig:corpus-competency-cooccurrence-heatmap} reveals logical correlations between competencies.
We observe two distinct clusters with strong positive correlations (scores > 0.3): one linking Social Awareness with Relationship Skills, and another connecting Self-Awareness, Self-Management, and Responsible Decision-Making. This indicates that our corpus effectively captures the synergistic nature of social skills, where understanding others is tied to interaction, and personal insight is linked to self-regulation and thoughtful action.

\textit{Social Skill Frequency}.
Figure~\ref{fig:corpus-top-10-social-skills} displays the frequency of the most common social skills within our corpus. Foundational abilities such as communication, empathy and compassion, and Perspective-taking are the most prevalent.

\textit{Scenario Context Distribution}.
The distribution of scenario contexts, visualized in Figure~\ref{fig:corpus-scenario-context-types}, is broadly balanced across key life domains. The most prominent categories are Family (31\%) and Workplace (27\%), with coverage of other areas like public and educational settings. This balance suggests the corpus provides contextually relevant scenarios for a range of real-world interpersonal challenges.

\subsection{\textbf{Expert Likert Ratings of Corpus}}
\label{appendix:expert-likert}
\textit{Tag Ratings}.
The expert panel rated the provided competence, skill, and context tags against the corresponding corpus content. Ratings of 1, 3, and 5 denoted incorrect, partially aligned, and directly aligned tags, respectively. As shown in Table~\ref{tab:corpus_quality_expert_ratings}, scenarios received the highest mean tag ratings, while theories and cases received lower ratings on the dimensions applicable to them. These results indicate that the sampled tags were generally viewed as aligned with the content.

\textit{Content Pedagogical Value}.
To assess content quality, the panel rated each knowledge item for Soundness (credibility and authenticity), Accessibility (clarity for a typical learner), and Pedagogy (potential usefulness for achieving a meaningful learning outcome).
As shown in Table~\ref{tab:corpus_quality_expert_ratings}, the results highlight the high pedagogical quality of the corpus across three dimensions. While scenario ratings consistently achieved the highest, the foundational theory scores are slightly lower. This indicates room for improvement in making abstract principles more engaging and instructionally impactful.
Overall, experts rated the sampled corpus items as credible, clear, and potentially useful, with scenarios receiving the strongest ratings for experiential learning.

\section{Evaluation on Scheduling and Tutoring}

\subsection{\textbf{Separation of Training and Evaluation Signals}}
\label{appendix:rubric-design}
SocialCoach uses three automatic components with non-overlapping roles. The frozen $E_\Delta$ updates only the synthetic planning state, $J_{\text{train}}$ provides pairwise pathway preferences for GRPO, and $J_{\text{eval}}$ independently rates held-out pathways without contributing to policy updates. Experts separately assess transition and pathway judgments, while tutoring outputs and user-study responses are evaluated outside the scheduling-training loop. These signals remain proxies and do not establish human learning gains.

\subsection{\textbf{Cross-Judge Robustness}}
\label{appendix:cross-judge-robustness}

The same frozen anonymous pathways are independently scored by GPT-5, Qwen3-235B, and DeepSeek-R1-0528 using the identical $1$--$5$ rubric and evidence bundle. For each dimension, we compute pairwise Spearman correlations over pathway-level scores, rather than over a small set of method means. Kendall's $W$ measures how consistently all three judges rank the same pathways. As shown in Table~\ref{tab:cross-judge-correlations}, the results show judge stablity cross difference models under the same rubric.

\begin{table*}[t]
\centering
\caption{Cross-judge agreement on pathway-level ratings. Pairwise columns report Spearman's $\rho$; Kendall's $W$ reports three-judge rank concordance.}
\vspace{-8pt}
\label{tab:cross-judge-correlations}
\small
\resizebox{\textwidth}{!}{%
\begin{tabular}{lcccc}
\toprule
\textbf{Metric}
& \textbf{GPT-5--Qwen3-235B $\rho$}
& \textbf{GPT-5--DeepSeek-R1-0528 $\rho$}
& \textbf{Qwen3-235B--DeepSeek-R1-0528 $\rho$}
& \textbf{Kendall's $W$} \\
\midrule
Goal Fit             & 0.72 & 0.68 & 0.75 & 0.71 \\
Personalization      & 0.65 & 0.61 & 0.69 & 0.64 \\
Pedagogical Value    & 0.78 & 0.74 & 0.80 & 0.77 \\
Sequence Coherence   & 0.69 & 0.66 & 0.72 & 0.68 \\
\bottomrule
\end{tabular}%
}
\vspace{-8pt}
\end{table*}

\subsection{\textbf{Alignment with Human Evaluation}}
\label{appendix:human-evaluation-alignment}

\paragraph{\textbf{Proficiency Delta Calibration.}}
The expert panel first calibrates the rubric, prompt, and ordered output anchors of $E_\Delta$ on a separate development set. Disagreements are used to refine the criteria and anchor definitions; the estimator and its anchor-to-delta mapping are then frozen. For independent validation, the panel labels a stratified sample of 100 transition cases that are disjoint from both calibration and policy training. The sample covers different target skills, initial proficiency levels, direct and cross-skill transitions, and novel versus repeated practice. Given the estimator inputs but not its output, each expert assigns an ordered expected-change label (\emph{no}, \emph{small}, or \emph{moderate}); the estimator's bounded numeric output is converted using the frozen anchors. As shown in Table~\ref{tab:delta-estimator-validation}, we report quadratic weighted $\kappa$ against the median expert label and ordinal Krippendorff's $\alpha$ across the three experts, with bootstrap 95\% confidence intervals.

\begin{table}[t]
  \centering
  \caption{Independent expert validation of the Expected Proficiency Delta Estimator on 100 held-out transition cases.}
  \vspace{-8pt}
  \label{tab:delta-estimator-validation}
  \small
  \begin{tabular}{lcc}
    \toprule
    \textbf{Comparison} & \textbf{Agreement} & \textbf{95\% CI} \\
    \midrule
    $E_\Delta$ vs.\ median expert label (weighted $\kappa$) & 0.56 & [0.38, 0.64] \\
    Three experts (Krippendorff's $\alpha$) & 0.61 & [0.44, 0.70] \\
    \bottomrule
  \end{tabular}
  \vspace{-8pt}
\end{table}

\paragraph{\textbf{Training-Judge Calibration.}}
We sample 100 pathway pairs from the frozen candidate groups of held-out profiles. The blinded expert panel independently selects A, B, or Tie from the learner goals and anonymous pathway evidence without seeing model identities, judge outputs, or numeric delta traces. Compared with the expert-majority label, $J_{\text{train}}$ achieves 74\% exact A/B/Tie agreement and Cohen's $\kappa=0.51$, indicating moderate alignment with expert pedagogical preferences.

\paragraph{\textbf{Expert Pathway Preference.}}
On a separate random sample of 100 held-out profiles, the blinded expert panel directly compares the complete pathways generated by SocialCoach$_{\text{Qwen3-8B}}$ and the strongest baseline, HA-RAG$_{\text{GPT-5}}$. Each pair uses the same learner profile, initial proficiency state, and corpus snapshot; pathway order is randomized, and model identities, automatic scores, and numeric delta traces are hidden. Panel members independently select A, B, or Tie. We aggregate the judgments by profile using the majority label and treat cases without a majority as ties. SocialCoach records 62 wins, 18 ties, and 20 losses, corresponding to a tie-adjusted preference rate of 68\% (profile-level bootstrap 95\% CI: [62\%, 74\%]). Inter-rater agreement on the original expert judgments is 0.61 according to nominal Krippendorff's $\alpha$.

\paragraph{\textbf{Pointwise-Evaluator Calibration.}}
\begin{table}[t]
  \centering
  \caption{Agreement between held-out pointwise evaluation and human experts.}
  \vspace{-10pt}
  \label{tab:correlations}
  \resizebox{\columnwidth}{!}{%
  \begin{tabular}{llcc}
    \toprule
    \textbf{Category} & \textbf{Metric} & \textbf{Correlation} & \textbf{p-value} \\
    \midrule
    \multirow{4}{*}{Learning Path}
      & Goal fit & 0.64 & \textit{p} < 0.01 \\
      & Personalization & 0.58 & \textit{p} < 0.01 \\
      & Pedagogical value & 0.62 & \textit{p} < 0.01 \\
      & Sequence coherence & 0.55 & \textit{p} < 0.01 \\
    \midrule
    \multirow{8}{*}{Tutoring Guidance}
      & Diagnosis          & 0.48 & \textit{p} < 0.01 \\
      & Attribution          & 0.39 & \textit{p} < 0.01 \\
      & Traceability       & 0.52 & \textit{p} < 0.01 \\
      & Specificity        & \underline{0.21} & \underline{\textit{p} > 0.05} \\
      & Actionability      & 0.45 & \textit{p} < 0.01 \\
      & Reflectiveness     & 0.36 & \textit{p} < 0.05 \\
      & Encouragement      & \underline{0.34} & \underline{\textit{p} > 0.05} \\
      & Personalization    & 0.41 & \textit{p} < 0.01 \\
    \bottomrule
  \end{tabular}%
  }
  \vspace{-10pt}
\end{table}

The blinded expert panel independently scores the same anonymous outputs as the held-out evaluator. For pathway quality, we use the predefined $1-5$ anchors and report rank correlation, weighted agreement, and confidence intervals. The results are shown in Table~\ref{tab:correlations}. The learning-path agreement results show moderate positive correlations across all four dimensions. The tutoring analysis likewise shows moderate agreement on several criteria; Specificity and Encouragement do not reach significance. Automatic scores remain proxy measurements rather than evidence of learning gains.



\subsection{Case Study of Learning Journey}
\label{appendix:case-studies}

\begin{figure}
    \centering
    \includegraphics[width=1.0\linewidth]{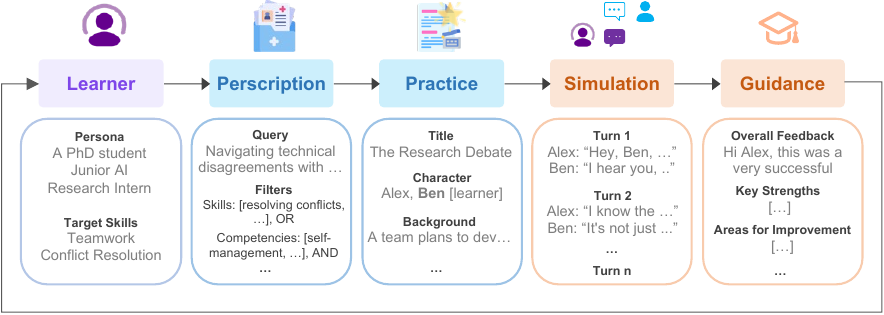}
    \vspace{-10pt}
    \caption{An illustrative learning journey case.}
  \vspace{-10pt}
    \label{fig:case-study}
\end{figure}

\begin{figure*}[!t]
  \centering
  \begin{subfigure}[b]{0.20\textwidth}
    \includegraphics[width=\linewidth]{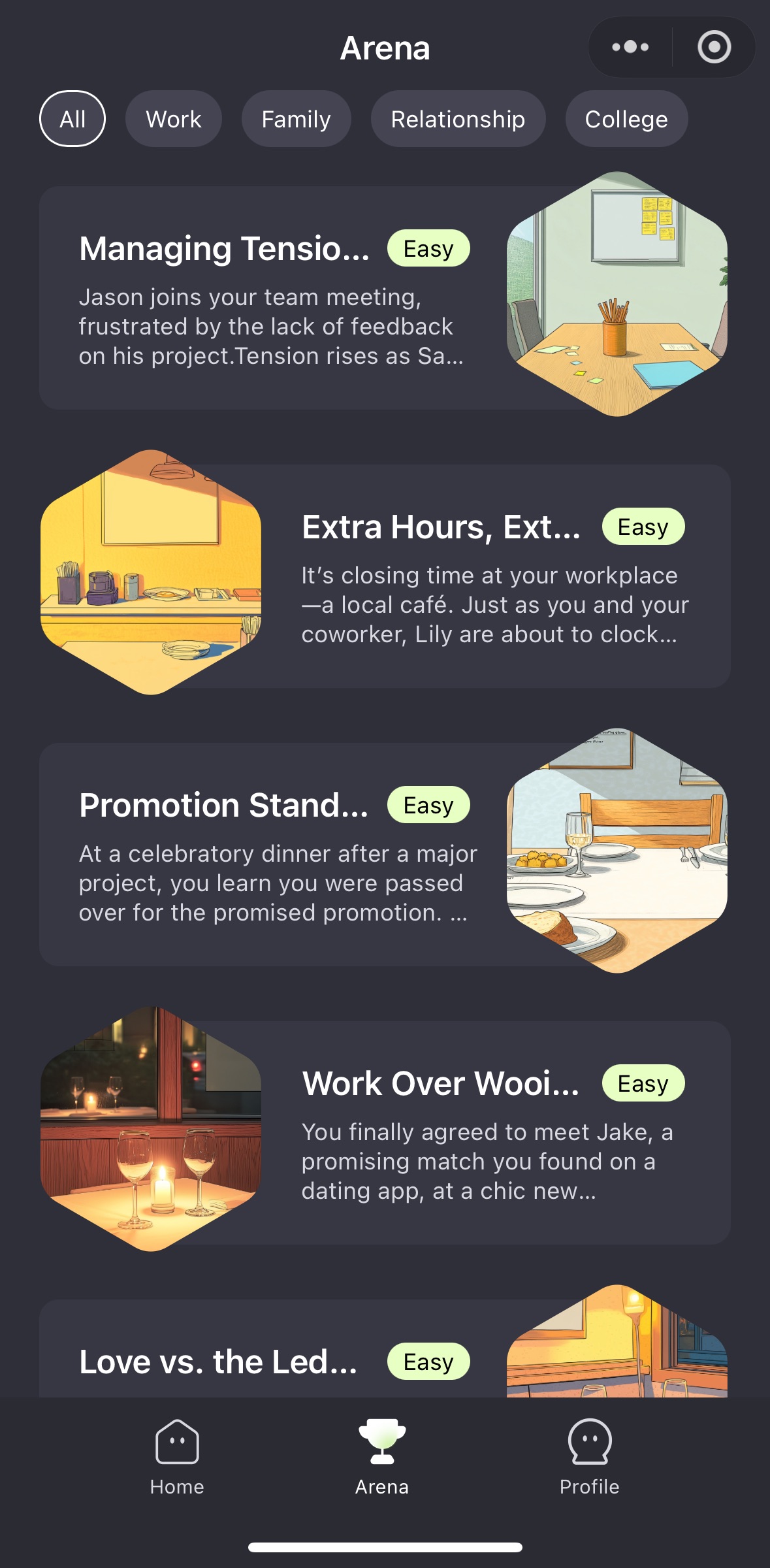}
    \label{fig:subfig1}
  \end{subfigure}
  \hfill
  \begin{subfigure}[b]{0.20\textwidth}
    \includegraphics[width=\linewidth]{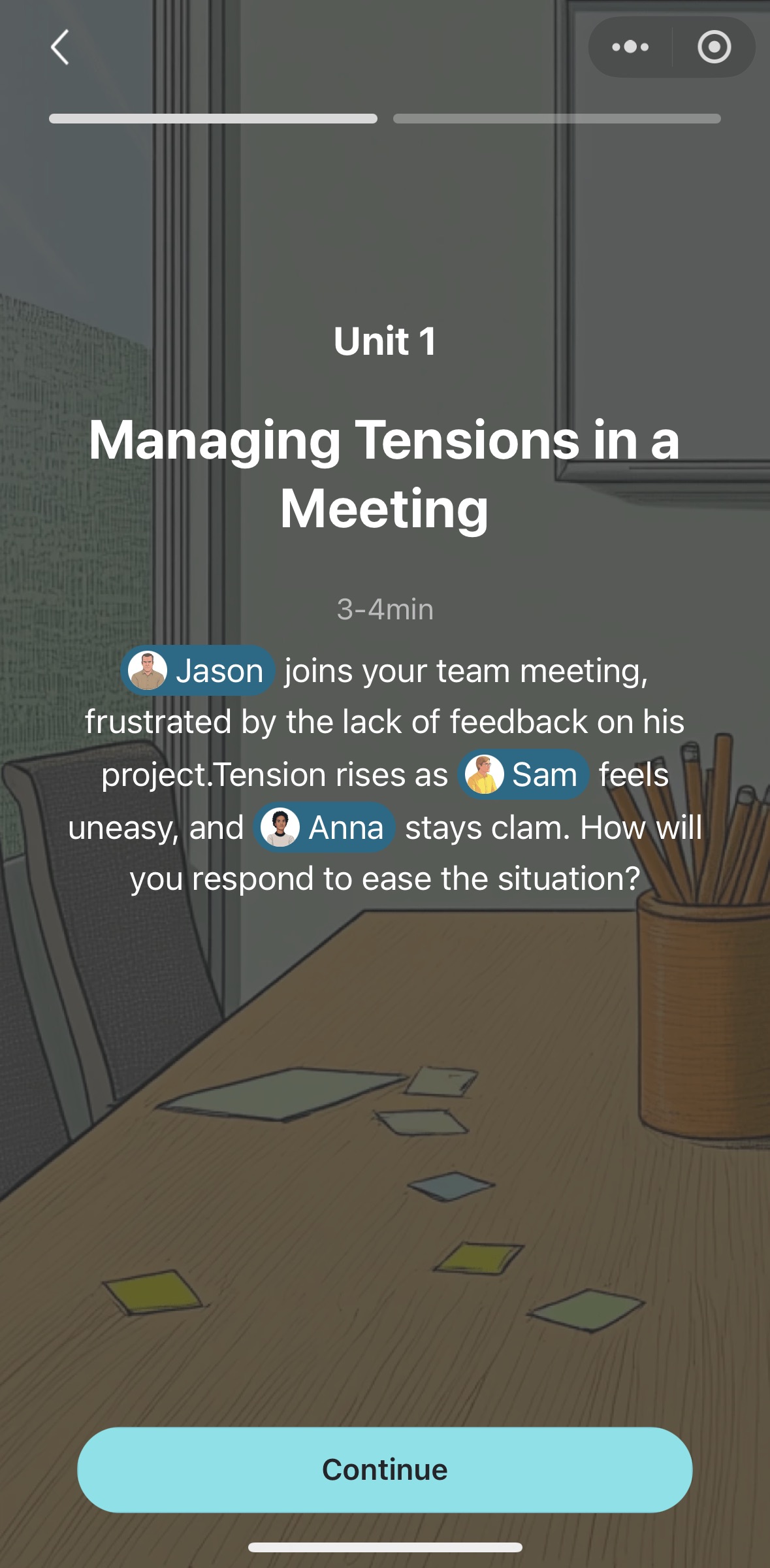}
    \label{fig:subfig2}
  \end{subfigure}
  \hfill
  \begin{subfigure}[b]{0.20\textwidth}
    \includegraphics[width=\linewidth]{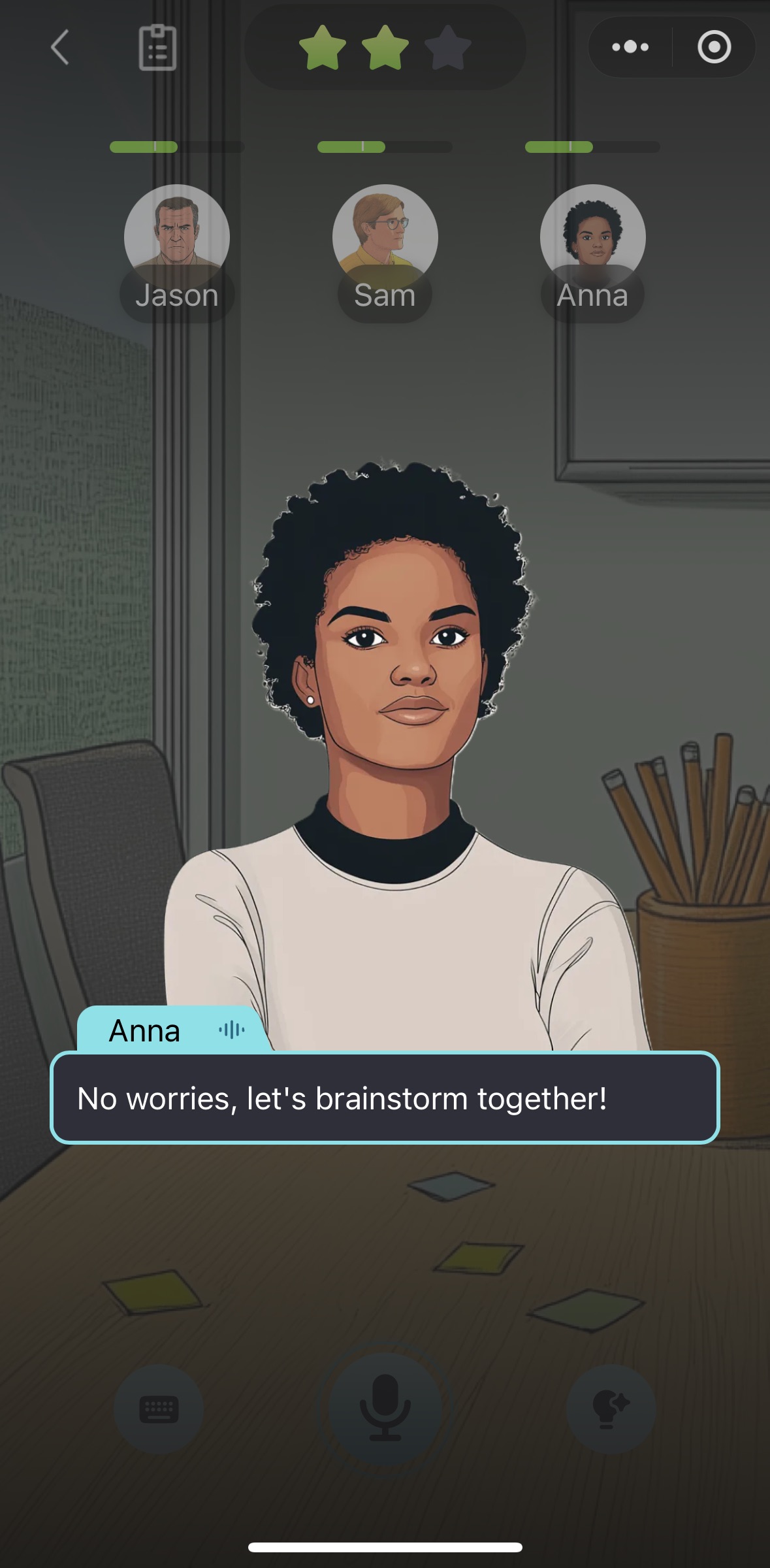}
    \label{fig:subfig3}
  \end{subfigure}
  \hfill
  \begin{subfigure}[b]{0.20\textwidth}
    \includegraphics[width=\linewidth]{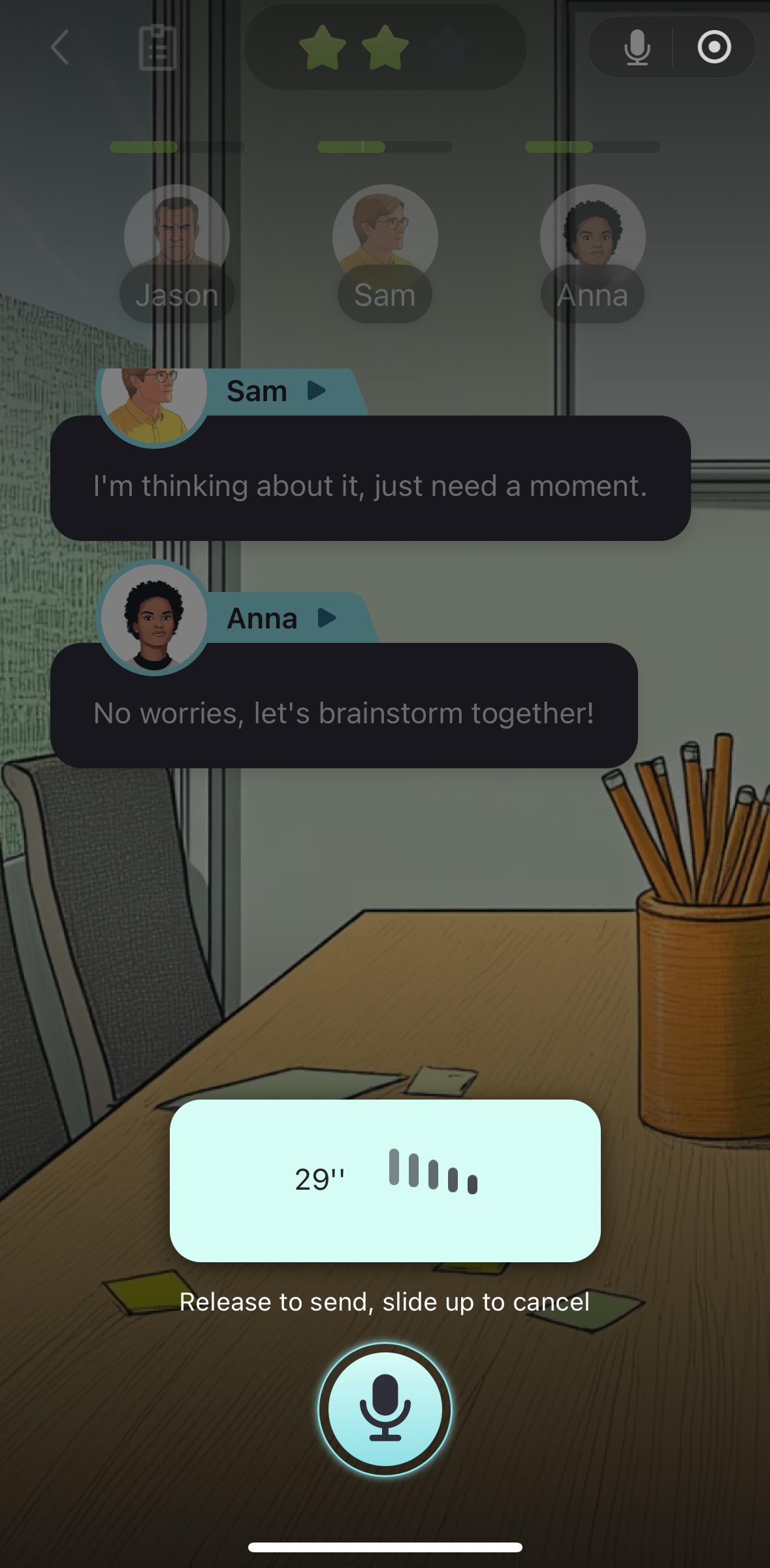}
    \label{fig:subfig4}
  \end{subfigure}
  \vspace{-10pt}
  \caption{Several interfaces of the internally deployed research platform.}
  \label{fig:demo}
  \vspace{-10pt}
\end{figure*}

\begin{figure*}
    \centering
    \includegraphics[width=0.96\linewidth]{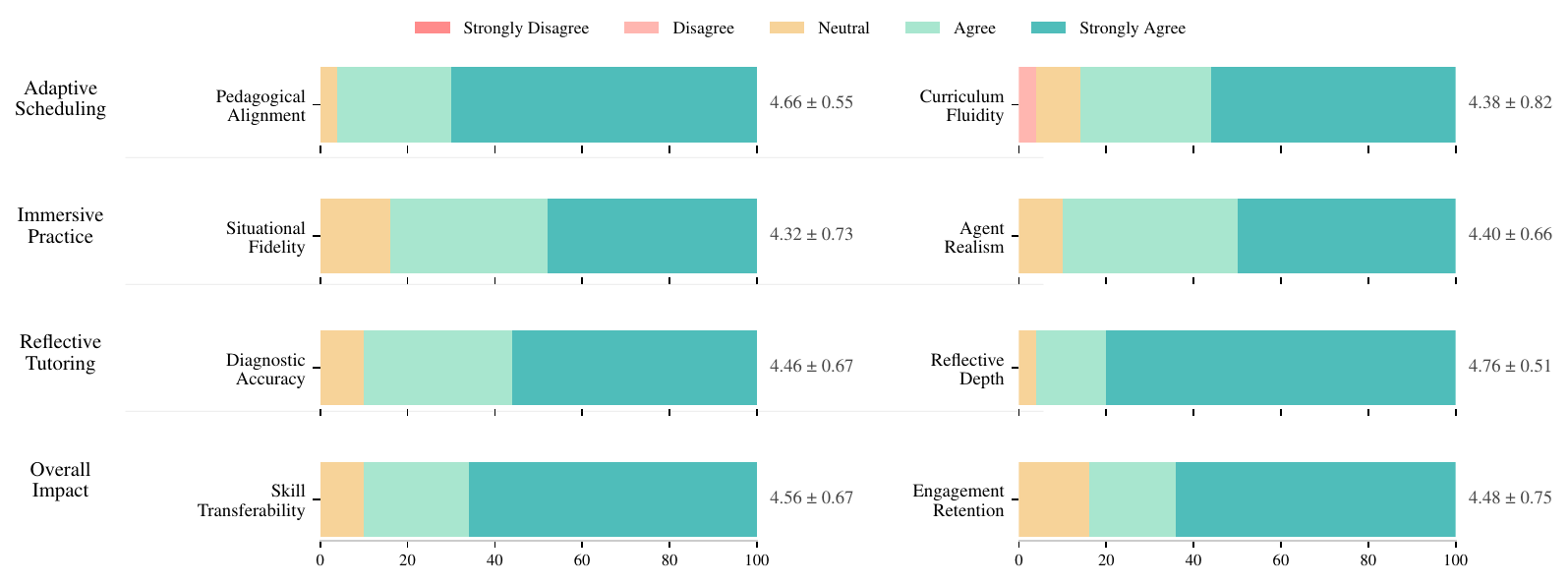}
    \vspace{-10pt}
    \caption{Self-reported questionnaire results from 50 participants. These outcomes measure perceived experience and intended application, not observed skill transfer.}
    \label{fig:user-study-results}
\end{figure*}

We illustrate a case study in Figure~\ref{fig:case-study}, featuring a learner whose persona is a junior AI research intern. This learner, while technically proficient, struggles with navigating professional disagreements and seeks to improve Teamwork and Conflict Resolution skills.

Based on this profile, the prescription agent generates a targeted query for a scenario about "navigating technical disagreements with peers," using filters to find content tagged with the Relationship Skills and Responsible Decision-Making competencies. The system retrieves a highly relevant scenario from the corpus titled "The Research Debate." In this practice, two researchers must resolve a conflict over choosing between a high-risk, innovative solution and a safer, established one under a tight deadline. Then, the adaptation agent assigns the learner the role of "Alex," the character advocating for the more ambitious, innovative approach.

In the simulation, the learner (as Alex) moves from defending his position to resolving the conflict. After validating his colleague's concerns about the project timeline, he proposes a concrete compromise: a "two-day spike" to test the new solution without
risking the schedule. They reached a consensus.

Upon completion, the system provides a detailed guidance report. The feedback praises the learner’s effective use of validation and integrative negotiation to reach a win-win outcome. It also identifies an area for growth: asking clarifying questions earlier to bypass the initial defensive exchange. Finally, the report poses reflective questions to help the learner consider how to establish shared goals. This process forms a complete learning loop.


\section{User Study via Deployed Platform}

\subsection{Description of Product Interfaces}
\label{appendix:product-demo}
We instantiated SocialCoach in an internally deployed research platform, whose key interfaces are presented in Figure~\ref{fig:demo}. After onboarding for initial profiling, the user journey begins in the practice arena, where learners are presented with a variety of scenarios like "Managing Tensions in a Meeting". SocialCoach's adaptive module dynamically schedules these scenarios to match the user's specific learning needs. Upon selecting a practice, the learner is briefed on the scenario's background, the characters involved (e.g., "Sam" and "Anna"), and the social objectives they need to achieve. The learner then enters an immersive simulation, engaging in a goal-driven conversation with LLM-powered agents who respond dynamically to their actions. This interactive dialogue serves as the basis for subsequent evidence-linked practice assessment and knowledge-grounded reflective tutoring. The gamified design of the platform illustrates how SocialCoach can be instantiated as an engaging learning experience that integrates adaptive practice with a complete pedagogical coaching loop.

\subsection{Analysis of User Study}
\label{appendix:user-study}

Results are shown in Figure~\ref{fig:user-study-results}; we summarize the main patterns below.

\textbf{Adaptive and Targeted Scheduling.} Participants highly valued the system's ability to align practice with individual needs. \textit{Pedagogical Alignment} was rated at 4.66 ± 0.55, suggesting that users perceived the system as matching social competencies to personal growth goals. \textit{Curriculum Fluidity} followed at 4.38 ± 0.82, with users noting that the adaptive pathways provided a logical progression through increasingly complex social challenges. One participant noted, ``The clear guidance made it easier to navigate addressing complex tasks and achieve goals.''

\textbf{Immersive Practice and Realism.} Participants perceived the simulation environment as realistic. \textit{Agent Realism} received a high rating of 4.40 ± 0.66, highlighting the perceived consistency of the AI agents' personas and their natural responses. \textit{Situational Fidelity} was rated at 4.32 ± 0.73, indicating that users felt the scenarios captured pressures similar to real-world social environments. As one participant stated, ``The interactions felt surprisingly real; I felt the same level of social pressure as I would in a physical workplace conflict.''

\textbf{Diagnostic and Reflective Guidance.} The reflective tutoring components achieved the highest marks in the study. \textit{Reflective Depth} stood out with a rating of 4.76 ± 0.51, suggesting that users found Socratic questioning useful for analyzing their strategies. The self-reported item labeled \textit{Diagnostic Accuracy} received 4.46 ± 0.67, indicating that participants perceived the diagnostic feedback as relevant. ``The feedback helped me discover the 'why' behind my behavior,'' a participant observed.

\textbf{Perceived Usefulness and Application Intent.} Participants reported perceived usefulness and confidence in applying practiced strategies. The questionnaire item labeled \textit{Skill Transferability} received 4.56 ± 0.67, reflecting intended application rather than observed skill transfer. The item labeled \textit{Engagement Retention} received 4.48 ± 0.75, reflecting continued-use intention rather than observed behavioral retention. One final interview revealed, ``[The platform] transformed my understanding of social intelligence into actionable skills I can use every day.''

\subsection{Pilot Learning Experience Study}
\label{appendix:learning-experience-study}

\paragraph{Design.}
We conducted a pilot learning experience study with 10 participants. After initial profiling, participants were assigned 1:1 to the adaptive or static condition using a computer-generated random permutation, yielding five participants per condition. SocialCoach generated a participant-specific ten-session practice path for every learner. The adaptive condition could revise the remaining path after each session using the learner's stated goals, updated synthetic planning state, and newly observed practice history; the static condition followed the initially personalized path without subsequent revision. Both conditions shared the same interface, scenario pool, simulator, tutoring and feedback modules, practice count, and time budget.

\paragraph{Measures.}
Rather than measuring stable trait change, we focused on practice pathway quality and immediate learning processes. After each session, participants rated their engagement with the scenario (1--5). The same expert panel, blinded to condition, later rated the pedagogical fit of each scheduled scenario (1--5). At the end of the study ($T_1$), participants wrote down the specific strategies they planned to apply next time. One week later ($T_2$), they were asked to recall those strategies without prompting. Session-level measures were first aggregated within each participant. Given the small sample ($n=5$ per condition), we report descriptive statistics only and do not conduct inferential tests.

\begin{table}[t]
  \centering
  \small
  \begin{tabular}{@{}lcc@{}}
    \toprule
    \textbf{Metric} & \textbf{Adaptive} & \textbf{Static} \\
    \midrule
    User engagement (1--5) & 4.4 $\pm$ 0.5 & 3.7 $\pm$ 0.6 \\
    Pedagogical fit (1--5) & 4.5 $\pm$ 0.9 & 3.8 $\pm$ 0.3 \\
    Avg. Strategy recall (immediate, $T_1$) & 2.8 $\pm$ 0.8 & 1.6 $\pm$ 0.5 \\
    Avg. Strategy recall (delayed, $T_2$) & 2.2 $\pm$ 0.7 & 0.6 $\pm$ 0.5 \\
    \bottomrule
  \end{tabular}
  \caption{Pilot learning experience study results. Values are descriptive means $\pm$ SD across participants.}
  \label{tab:learning-experience-results}
  \vspace{-10pt}
\end{table}

\paragraph{Interpretation.}
In this small pilot, the adaptive group showed numerically higher means across all four outcomes. The between-group differences were 0.7 points for both engagement and expert-rated pedagogical fit, 1.2 recalled strategies at $T_1$, and 1.4 recalled strategies at $T_2$. The common direction across participant-rated engagement, blinded expert fit, and both recall measurements is consistent with the hypothesis that session-by-session path revision can maintain alignment with learner goals and practice history; the relatively larger delayed-recall difference makes retention a useful target for a larger confirmatory study. With only five participants per condition and no inferential tests, however, these patterns are exploratory and do not establish population-level or causal effects.

\section{Further Discussion}

\subsection{Modeling Scope and Limitations}
Longer-term proficiency changes, including forgetting, are not explicitly modeled; when an updated learner profile becomes available, its proficiency state can initialize a subsequent scheduling episode.

\subsection{Safety, Fairness, and Cultural Context}
SocialCoach is intended for low-stakes practice and reflection, not clinical assessment, employment decisions, or other high-stakes judgments. Numeric initial proficiency and delta transitions are used only inside the synthetic cold-start environment: starting scores are not inferred from demographic attributes, and predicted deltas are not presented as assessments of real users. Maximum Marginal Relevance (MMR)~\cite{profile-sigir-1998-mmr-diversity-reranking} and one-to-one matching with the Hungarian algorithm~\cite{profile-nrlq-1955-hungarian-assignment-problem} are used to sample diverse profile combinations, but these procedures are sampling steps rather than guarantees of demographic fairness or stereotype avoidance. The system may still reproduce assumptions present in its source material or model outputs. Because the corpus and profile sources are primarily English-language and culturally situated, we do not claim cross-cultural validity. SocialCoach currently focuses on text-based, language-mediated practice and does not observe non-verbal or embodied cues such as prosody, gaze, gesture, facial expression, or physiological responses. Accordingly, the reported evidence should not be interpreted as covering embodied social competence or transfer to face-to-face settings. Scenarios and guidance should therefore be interpreted as context-conditioned practice prompts, and the current expert review should not be read as a comprehensive fairness or cultural audit.

\subsection{Data Provenance and Licensing}
The corpus was constructed from 200 books accessed through Open Library and transformed into structured theories ($K_t$), cases ($K_c$), and scenarios ($K_s$). We retain source metadata and provenance for derived items, while the underlying source texts and local download files are not part of the paper artifact. The complete experiment corpus is maintained separately from the released illustrative examples. We do not redistribute the source books or the complete experimental corpus. We do not make a blanket copyright or fair-use determination in this paper: source-specific licenses, access terms, and institutional review govern redistribution. Any artifact release is therefore limited to code, schemas, manifests, and derived examples permitted by those terms.

\end{document}